\documentclass{article}
\usepackage[a4paper,
            bindingoffset=0.2in,
            left=0.6in,
            right=0.6in,
            top=1in,
            bottom=1in,
            footskip=.25in]{geometry}
\usepackage[utf8]{inputenc}
\usepackage{jcappub}
\usepackage{natbib}
\setcitestyle{square,numbers}
\usepackage{graphicx}
\usepackage{xcolor}
\usepackage{float}
\usepackage{multirow,array,longtable}
\usepackage{arydshln} % Dashed lines in array. siempre despues de 'array'
\usepackage[makeroom]{cancel}
\usepackage{amsfonts}%,mathalfa,amssymb,amsmath}

\renewcommand\eqref[1]{Eq.~(\ref{#1})}

\newcommand\figref[1]{Fig.~\ref{#1}}

\newcommand\secref[1]{Section~\ref{#1}}

\newcommand{\be}{\begin{equation}}
\newcommand{\ee}{\end{equation}}
\newcommand{\bear}{\begin{eqnarray}}
\newcommand{\eear}{\end{eqnarray}}
\newcommand{\nn}{\nonumber}

\def\1loop{one-loop}
\def\amp{\mathcal{A}}
\newcommand{\mL}{\mathcal{L}}

\def\mh{m_{h}}
\def\mw{m_{W}}
\def\mz{m_{Z}}

\def\vev{{\it v}}
\def\cb{c_\beta}
\def\sb{s_\beta}
\def\tanb{t_\beta}
\def\tb{\tan \beta}

\newcommand{\sba}{s_{\beta-\alpha}}
\newcommand{\cba}{c_{\beta-\alpha}}
\def\mHp{m_{H^\pm}}
\def\mHeavy{m_H}
\def\mA{m_A}
\def\lahHpHm{\lambda_{hH^+H^-}}
\def\lahhH{\lambda_{hhH}}
\def\lahhh{\lambda_{hhh}}

\def\lahhhh{\lambda_{hhhh}}

\def\mbarc{\frac{m_{12}^2}{s_\beta c_\beta}}%{\bar{m}^2}
\def\mbarh{\frac{m_{12}^2}{\mh^2s_\beta c_\beta}}%{\frac{\bar{m}^2}{\mh^2}}
\title{Matching 2HDM to HEFT in triple Higgs boson production from WBF}
\author{L. Barreiro,}
\author{D. Domenech,}
\author{D. Garcia-Tejada,}
\author{M. Herrero, }
\author{S. Salazar}
\affiliation{Departamento de F\'{\i}sica Te\'orica and Instituto de F\'{\i}sica Te\'orica, IFT-UAM/CSIC,\\
Universidad Aut\'onoma de Madrid, Cantoblanco, 28049 Madrid, Spain}
\emailAdd{lucia.barreiro@estudiante.uam.es}
\emailAdd{daniel.domenech@uam.es}
\emailAdd{diego.garciat@estudiante.uam.es}
\emailAdd{maria.herrero@uam.es}
\emailAdd{sebastian.salazar@estudiante.uam.es}

\abstract{In this work we  compare the Higgs Effective Field Theory  (HEFT) predictions with the Two Higgs Doublet Model (2HDM) predictions for   triple Higgs production via $WW \to hhh$ scattering,  under the hypothesis of very heavy Higgs bosons  $H$,  $A$,  and $H^\pm$,  and assuming the lightest one $h$ has the observed  mass in the experiments,  $m_h \simeq 125$ GeV.  By requiring the matching of the two corresponding  scattering amplitudes for this $WW \to hhh$ process in the limit of very heavy mass for these bosons,  we find the solutions for the leading order HEFT coefficients.  Concretely,  we determine the coefficients $a$, $b$, $c$,  $\kappa_3$ and $\kappa_4$  defining the effective interactions $WWh$,  $WWhh$, $WWhhh$, $hhh$ and $hhhh$,  respectively,  in terms of the 2HDM input parameters,  $\cba$,  $\tan \beta$ and $m_{12}$.  We will also compare these solutions with the corresponding solutions found for the case of double Higgs production.  The solutions presented here for the matched coefficients summarize the non-decoupling  effects from the heavy modes,  $H$, $A$ and $H^\pm$ in the scattering process 
$WW \to hhh$ and provide the most relevant contributions to the cross section $\sigma(WW \to hhh)$ that do not decrease as inverse powers of the heavy masses.  In the last part,  we will also discuss the numerical predictions of the cross section $\sigma(WW \to hhh)$ in terms of the HEFT-matched coefficients.  The focus of this discussion is set on the big gap between the HEFT-matched prediction and the SM prediction due to the non-decoupling behaviour occurring whenever misalignment is assumed in the 2HDM, i.e.  for $\cba \neq 0$.  This big gap could serve as potential signals of the 2HDM heavy modes,  $H$, $A$ and $H^\pm$,  at future high energy colliders,  even if they are too heavy and cannot be produced at these colliders.}

\begin{document}
\begin{flushright}
   \today\\
	IFT-UAM/CSIC-26-147
\end{flushright}
\maketitle
%%%%%%%%%%%%%%%%%%%%%%%%%%%%%%%
\section{Introduction}
\label{section-intro}

One of the most appealing features of the use of Effective Field Theories to describe Beyond Standard Model (BSM) Physics  is that they provide model independent predictions for the scattering amplitudes at low energies which can be tested at colliders.   The information of the new Physics is encoded in the values of the effective coefficients (usually called Wilson coefficients) which can be computed from the underlying full theory by integrating out the heavy degrees of freedom,  and keeping at low energies just the light degrees of freedom which are identified with the Standard Model (SM) particle content.  The Effective Field Theory  (EFT)  then contains the relevant set of BSM effective interactions among the SM fields that can be explored at colliders by means of observables like cross sections, distributions,  decay rates,  etc.   These observables,  by construction,  are defined in terms of SM particles in the external legs.  In particular,  the new effective interactions among SM bosons  (Higgs and gauge bosons)  may contain the so-called anomalous effective couplings  like,  for instance,  triple and quartic gauge boson couplings, triple and quartic Higgs couplings,  and the couplings of gauge bosons to multiple Higgs bosons,  most of them  being trackeable at colliders (for a status-review of searches at colliders of these anomalous effective couplings see \cite{ParticleDataGroup:2026mpi}).  Concretely,  we will focus here on the effective interactions of  electroweak gauge bosons $V=W,Z$  to multiple Higgs bosons and on the Higgs self-interactions,  and choose as EFT the case of Higgs Effective Field Theory (HEFT).   In contrast to the more popular Standard Model Effective Field Theory (SMEFT) the HEFT is better suited for scenarios where the underlying ultraviolet (UV) theory could contain large couplings which is the main focus of our interest here. 

Among the various candidates for the new physics BSM,  the two Higgs doublet model (2HDM) is probably one of the most studied ones,  since this is the simplest extension of the SM containing two Higgs doublets instead of just one like in the SM (for a review of the 2HDM see for instance~\cite{HHG,Branco:2011iw}).  The physical Higgs sector of the 2HDM contains five Higgs bosons,  two CP-even,  $h$ and $ H$,  one CP-odd,  $A$,  and two charged ones,  $H^{\pm}$.   The lightest one $h$ is usually  identified with the Higgs boson observed experimentally with a mass value of $m_h \sim 125$ GeV~\cite{ATLAS:2012yve,CMS:2012qbp,ATLAS:2015yey} and the others are searched for intensely at colliders but no-signal of their existence have been found so far.  As a result of these failed searches,  the input parameters of the 2HDM are presently constrained.  In particular,  the masses of the new Higgs bosons, are pushed to be above several hundred of GeV in most of the experimental searches (for a status-review see  ~\cite{ParticleDataGroup:2026mpi}).   Here we will assume that these new Higgs bosons,  $H$, $A$ and $H^{\pm}$  get heavy masses,  generically called $m_{\rm heavy}$,  with numerical values  above the typical EW scale given by $v=246$ GeV .  Our general  objective is then to use the proper EFT approach to describe the low energy effects of the heavy BSM Higgs bosons scenario of the 2HDM within the non-decoupling regime, i.e.  where these low energy effects are not suppressed by inverse powers of $m_{\rm heavy}$.

In this work,  we use the HEFT as the proper tool to capture all the potential non-decoupling effects from the heavy BSM Higgs bosons, $H$, $A$ and $H^\pm$  in the case where the UV underlying theory is the 2HDM.  This heavy mass assumption for the new Higgs bosons may imply in turn large triple and quartic couplings  in the 2HDM Higgs sector and these are precisely the responsible for the appearance of non-decoupling effects at low energies.  In contrast,  assuming the SMEFT as the low energy EFT of the 2HDM always leads to a decoupling scenario,  with all new physics effects being suppressed by inverse powers of the heavy Higgs boson masses.  Previous studies of the low energy EFT for the 2HDM heavy scenario within the decoupling regime have been done mostly in the context of the SMEFT ~\cite{Belusca-Maito:2016dqe,Dawson:2022cmu,Banerjee:2022thk,Banerjee:2023qbg}.  In  ~\cite{Dawson:2023ebe} both the SMEFT and HEFT cases were considered in the decoupling regime.  The first studies of the non-decoupling scenario for the 2HDM within the HEFT were done in \cite{Arco:2023sac}.  They used the technique of matching the predictions for the scattering amplitudes of the two theories,  2HDM and HEFT,  for a series of specific processes with on-shell SM particles in the external legs.  In particular,  they  solved   the tree level matching for the following cases,  Vector Boson Fusion (VBF) processes like $WW \to h$, $ZZ \to h$,  $WW \to hh$,  $ZZ \to hh$ and also elastic scattering $hh \to hh$.   By a detailed analysis of all the Feynman diagrams contributing to these scattering amplitudes in both the HEFT and the 2HDM,   and  performing a large $m_{\rm heavy}$ expansion in the 2HDM case,  they found the solutions to the matching,  providing  the values of the HEFT coefficients in terms of the input parameters of the 2HDM.  Concretely they found the values of the Leading Order (LO) HEFT parameters describing the following effective interactions: $VVh$ (parameter $a$,  also called $\kappa_V$ in the literature) ,  $VVhh$ (parameter $b$,  also called $\kappa_{2V}$ in the literature),  $hhh$ 
(parameter $\kappa_3$,  also called $\kappa_\lambda$ in the literature)  and $hhhh$ (parameter $\kappa_4$), and they were presented in terms of the input parameters of the 2HDM.   These non-decoupling solutions were confirmed later by other works \cite{Buchalla:2023hqk,  Dawson:2023oce}.  In \cite{Buchalla:2023hqk} they used the different functional method approach to integrate out the heavy Higgs bosons and discussed how to reach the two different decoupling and non-decoupling regimes.  The differences in the results of these HEFT coefficients for the  different  decoupling and non-decoupling  scenarios of the 2HDM  were also discussed in \cite{Dawson:2023oce}.  As a general conclusion in all these works \cite{Arco:2023sac,  Buchalla:2023hqk,  Dawson:2023oce},  the 2HDM parameter $\cos(\beta -\alpha)$ seems to play a very relevant role in getting decoupling or non-decoupling behaviour of the heavy Higgs bosons at low energies.  If one assumes this $\cos(\beta -\alpha)$ parameter to be a free input parameter with numerical values just limited by present experimental constraints (given approximately by
$|\cos(\beta-\alpha)|<0.35$)  one gets non-decoupling scenarios where the low energy effects are not suppressed by inverse powers of $m_{\rm heavy}$.   The non-decoupling effects can be summarized by the values of $a$,  $b$, $\kappa_3$ and $\kappa_4$ of ref. \cite{Arco:2023sac}.  More specifically,  by the values of $\Delta a$, $\Delta b$, $\Delta \kappa_3$ and $\Delta \kappa_4$ that are defined as the separation with respect to their SM reference value of 1,  i.e.  $\Delta {\rm coeff}  \equiv 1- {\rm coeff}$.  Notice that all these non-decoupling effects in $\Delta {\rm coeff}$  are of 
${\cal O}(\cos^2(\beta -\alpha))$.  In the alignment limit,  defined by  
$\cos(\beta -\alpha) \to 0$ all these non-decoupling  effects vanish,  as expected,  leading to the SM reference values $a, b,\kappa_3, \kappa_4  \to 1$.  But this is not so if one departs from this alignment limit.  For instance, in the so-called quasi-alignment limit with small but non-zero values,  say  
$|\cos(\beta-\alpha)| < 0.35$,   the size of the non-decoupling  effects are still large.  On the contrary,  if   $\cos(\beta -\alpha)$ is considered not as a free input parameter but instead it is a derived parameter in these heavy Higgs scenarios of the 2HDM,  then the resulting EFT manifest a decoupling behaviour of the integrated heavy modes,  i.e with the low energy effects being suppressed by inverse powers of $m_{\rm heavy}$ , and this mainly being a consequence of the 'set in'  behaviour of the derived 
$\cos(\beta -\alpha) \sim {\cal O}(v^2/m^2_{\rm heavy})$.  

Our main objective here is to continue the matching procedure  of the 2HDM to the HEFT in the non-decoupling regime of the 2HDM heavy Higgs bosons scenario by using a similar amplitudes matching method as in  \cite{Arco:2023sac} but now for a different scattering process  with much interest for phenomenology at future high energy colliders.  Concretely we will solve here for the first time the matching in the case of triple Higgs production from  the VBF  of two charged electroweak gauge bosons,  namely,  $WW \to hhh$.  This will allow us to fulfill two objectives.  First to perform a check with this new scattering process of the previous solutions found in  \cite{Arco:2023sac} for $a$, $b$ and $\kappa_3$ and $\kappa_4$.   Second, and more importantly,  it will also allow us to get, using diagrammatic methods,  the value of the new effective interaction of two $W$'s to three Higgs bosons $WWhhh$ given  by the $c$ parameter.  In contrast to the previous HEFT  parameters $a$,  $b$,  $\kappa_3$ and $\kappa_4$ whose potential departures from  1 (their  SM reference value) define deviations to the SM couplings,  the value of $c$ defines a totally new contact interaction which is not present in the SM.  Hence,  the potential impact of this non-vanishing  $c$ parameter for future searches of BSM Higgs Physics at colliders is of most interest. 

The paper is organized as follows.  \secref{HEFT-2HDM} provides a brief review of the basic ingredients  of the HEFT and the 2HDM which are needed for this work.  The analytical computation of the new $WW \to hhh$ amplitudes  within  HEFT and 2HDM is presented  in \secref{section-amplitudes}.  The matching procedure among these scattering amplitudes is described in \secref{section-matching}.  The way to the solution for the HEFT coefficients in terms of the input 2HDM parameters that summarizes  the non-decoupling effects in the heavy mass regime is also included in this section.  The numerical analysis of the previous results are presented in \secref{section-numerical}.   Finally, we conclude in \secref{section-conclus}. 
%The details of the Feynman rules and the one-loop functions are given in the appendices.

%%%%%%%%%%%%%%%%%%%%%%%%%%%%%%%
\section{Theoretical framework: HEFT and 2HDM}
\label{HEFT-2HDM}

Here we describe in short the theoretical framework for both the HEFT and the 2HDM that is needed for the present computation of the scattering amplitudes for triple Higgs production from WBF.  We work at the tree level approximation for these scattering amplitudes and focus on the bosonic electroweak  sector of these two theories which is the relevant one to describe the  process of our interest here,  $WW \to hhh$.

%%%%%%%%%%%%%%%%%%%%%%%%%%%%%%%
\subsection{The basics of HEFT}

%%%%
We summarize first the basics of the bosonic  HEFT Lagrangian that is relevant for this work.   For the tree level computation of the $WW \to hhh$ scattering amplitude it is enough to work with the leading order (LO) HEFT Lagrangian,  i.e,  with chiral dimension 2.   The setting of the LO-HEFT Lagrangian in an arbitrary $R_\xi$  gauge was given in ~\cite{paperHdecays,Herrero:2021iqt,Herrero:2022krh} which we follow closely here.  This HEFT  Lagrangian  is written in terms of the light bosonic fields, namely,  the EW gauge bosons, $B_\mu$ and $W^a_\mu$ ($a=1,2,3$), their corresponding Goldstone Bosons (GBs)  $\omega^a$ ($a=1,2,3$), and the Higgs boson $h$.  The HEFT Lagrangian is invariant under EW gauge,  $SU(2)_L\times U(1)_Y$,  transformations.  The pure scalar sector of the HEFT has an additional global invariance named usually  EW chiral $SU(2)_L \times SU(2)_R$ symmetry.  The LO-HEFT Lagrangian in a covariant $R_\xi$ gauge is given by:
     \begin{eqnarray}
      {\mL}_{\rm HEFT}   &=& \frac{v^2}{4}  \left(1 + 2  a \frac{h}{v} +b \frac{h^2}{v^2}+c  \frac{h^3}{v^3}+... \right)
        \text{Tr}[(D_{\mu} U)^{\dagger} D^{\mu} U] + \frac12 \partial_{\mu} h  \partial^{\mu} h - V(h) \nn \\
       && - \frac{1}{2 g^2} \text{Tr}[\hat{W}_{\mu \nu} \hat{W}^{\mu \nu}] - \frac{1}{2 g'^2} \text{Tr}[\hat{B}_{\mu \nu} \hat{B}^{\mu \nu}] + {\mL}_{GF} + {\mL}_{FP},
        \label{LO-HEFT}
    \end{eqnarray}
One very important feature of the HEFT (in contrast to the SMEFT) is the properties of the Higgs field $h$ being inferred from  its singlet character under all relevant symmetries.  This singlet character of $h$ is also in contrast 
to the SM case where $h$ is introduced instead as a component of the usual complex doublet field
$\Phi$,  containing 4 scalar fields.  Due to this singlet nature of $h$ within the HEFT,  the first term in \eqref{LO-HEFT} contains a polynomial function in powers of $(h/v)$.  For the present work, we only need to keep terms in this function up to ${\cal O}(h^3)$. 

The $U$ field is a $2\times2$ matrix
\begin{equation}
        U \, = \, \exp \left( i \frac{{\omega_i} {\tau_i}}{v} \right),
    \end{equation}  
   that contains the three GB fields ${\omega}_i$ ($i=1,2,3$) in a non-linear representation of the $SU(2)$ symmetry group and 
   %$t_i (i=1,2,3)= \tau_i/2$ are the three generators of $SU(2)$,  
  % with 
   $\tau_i$ ($i=1,2,3$) are the three Pauli matrices.

   The EW covariant derivative of this $U$ field 
   is defined as:
   \begin{equation}
        D_{\mu} U \, = \, \partial_{\mu} U - i \hat{W}_{\mu} U + i U \hat{B}_{\mu},
    \end{equation} 
    that contains the EW gauge fields,  $\hat{W}_{\mu} = \frac{g}{2} W^i_{\mu} \tau^i$ and $\hat{B}_{\mu} = \frac{g'}{2} B_{\mu} \tau^3$ and the EW gauge couplings $g$ and $g'$.  The corresponding EW field strength tensors are given by\footnote{Notice that we are using here a different sign convention in the covariant derivative and the field strength tensor than in ~\cite{paperHdecays,Herrero:2021iqt,Herrero:2022krh,  Arco:2023sac}.  This does not affect to any result in this work.}:
       \begin{equation}
        \hat{W}_{\mu \nu} \, = \, \partial_{\mu} \hat{W}_{\nu} - \partial_{\nu} \hat{W}_{\mu} - i [\hat{W}_{\mu}, \hat{W}_{\nu}], \hspace{8mm} \hat{B}_{\mu \nu} \, = \, \partial_{\mu} \hat{B}_{\nu} - \partial_{\nu} \hat{B}_{\mu}.
    \end{equation}
    
  \begin{figure}[!t]
    \centering
        \includegraphics[scale=0.3]{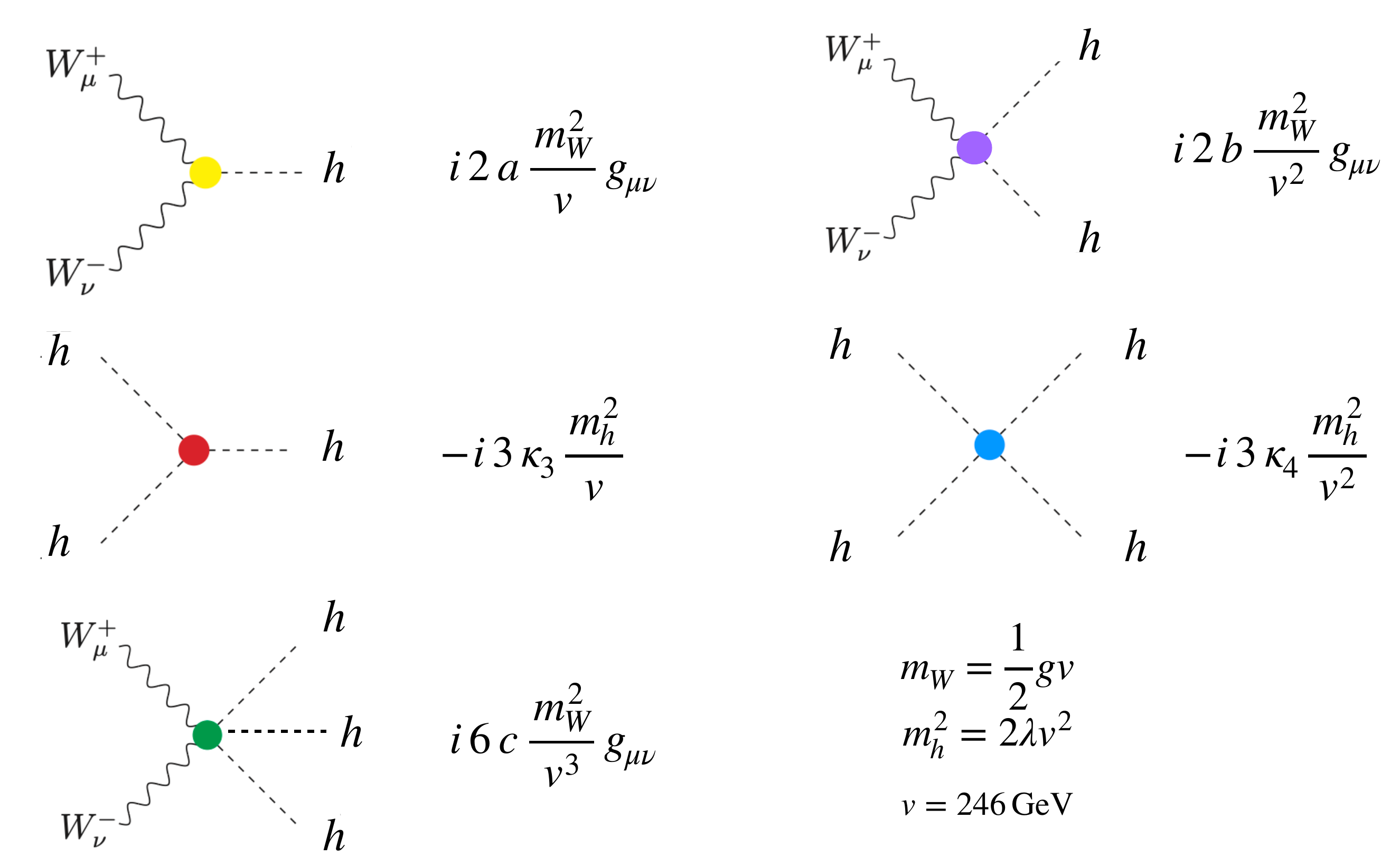}
    \caption{Relevant Feynman Rules for the LO-HEFT in the unitary gauge.  They are written here in terms of the physical on-shell parameters, $m_h$,  $m_W$ ,  $v= 246$ GeV  and the LO-HEFT coefficients $a$, $b$, $c$, $\kappa_3$ and $\kappa_4$.  The FRs of the SM can be recovered from the FRs of the HEFT shown here,  with  the specific setting $a=b=\kappa_3=\kappa_4=1$ and $c=0$.  }
    \label{FRs-HEFT}
    \end{figure}
      
    The physical gauge fields are then given by:
\be
W_{\mu}^\pm = \frac{1}{\sqrt{2}}(W_{\mu}^1 \mp i W_{\mu}^2) \,,\quad
Z_{\mu} = c_W W_{\mu}^3 - s_W B_{\mu} \,,\quad
A_{\mu} = s_W W_{\mu}^3 + c_W B_{\mu} \,,
\label{eq-gaugetophys}
\ee
where we use the short notation $s_W=\sin \theta_W$ and $c_W=\cos \theta_W$,  with $\theta_W$ the weak angle.

The function $V(h)$ in \eqref{LO-HEFT} is the Higgs potential within the HEFT, which includes the triple and quartic  Higgs self-interactions:
    \begin{equation}
        V(h) \, = \, \frac12 m_h^2 h^2 +\kappa_3 \lambda v h^3 + \kappa_4 \frac{\lambda}{4} h^4.
    \label{VHEFT}    
    \end{equation}
The relations among couplings  and masses within the LO-HEFT are as in the SM:   $m_h^2  = 2 \lambda v^2$,  $\mw=g v /2$ and $\mz=\mw/c_W$.  
Here we will identify the $h$ state with the Higgs boson discovered at the LHC with a mass $m_h= 125$ GeV~\cite{ATLAS:2012yve,CMS:2012qbp,ATLAS:2015yey}. 
 The vacuum expectation value $v$ can be written  in terms of  the Fermi coupling constant $G_F$ by $v=(\frac{1}{\sqrt{2} G_F})^{1/2} =246$ GeV. 

The functions  ${\mL}_{GF}$ and ${\mL}_{FP}$  in \eqref{LO-HEFT} are the gauge fixing  and Faddeev-Popov ${\mL}_{FP}$ terms,  respectively, whose explicit expressions in the general $R_\xi$ covariant gauges can be found in  ~\cite{paperHdecays,Herrero:2021iqt,Herrero:2022krh} together with more definitions and specifications within the HEFT. 
%(for a more complete list of effective operators within the HEFT see also \cite{Brivio:2013pma}).  
For the present computation of the $WW \to hhh$ scattering amplitude we will work instead within the unitary gauge,  i.e.  where there are no GBs  ($U=1$), and the gauge propagators for the EW fields are fixed to the unitary gauge.  In any case we have checked explicitly the  gauge invariance of the resulting amplitude.  This gauge invariance is expected since the HEFT is a $SU(2)_L \times U(1)_Y$ gauge invariant EFT.  The relevant Feynman rules for the LO-HEFT in the unitary gauge are summarized in Fig. \ref{FRs-HEFT}.

%%%%

Finally, just to recall  the way the SM is embedded within the HEFT.  It is clear that this embedding cannot be done at the Lagrangian level,  since as we already said  within the SM the Higgs field is introduced into a doublet whereas in the HEFT it is a singlet.  Thus, the Lagrangian themselves are not directly comparable.  The Feynman rules for the couplings with scalar fields ($h$ and the GBs)   are also different in the HEFT and the SM due to the non-linear parametrization for the GBs used in the HEFT.  Thus,  the comparison between the SM and the HEFT should be done via their predictions for the observables instead of via their corresponding Lagrangians.  Specifically,  in order to reach the SM predictions from the LO-HEFT predictions one has to fix the LO-HEFT coefficients to the following particular values: $a=1$,  $b=1$,  $\kappa_3=1$,  $\kappa_4=1$ and $c=0$.  Alternatively,  if we write the coefficients $a$, $b$,  $\kappa_3$ and $\kappa_4$ in terms of the corresponding $\Delta$'s,  which are  defined as $a=1-\Delta a$,  $b=1-\Delta b$, $\kappa_3= 1-\Delta \kappa_3$ and $\kappa_4= 1-\Delta \kappa_4$,  then the SM predictions can be obtained from the HEFT predictions by setting,  $\Delta a =0$,  $\Delta b =0$,  $\Delta \kappa_3=0$,  $\Delta \kappa_4=0$,  in addition to setting $c=0$ .  

%We have summarized  the Feynman rules of the HEFT that are relevant for the present computation in \tabref{relevant-FR} of \appref{App-Frules} 
%(for a full set see \cite{paperHdecays,Herrero:2021iqt,Herrero:2022krh}). The corresponding Feynman rules of the SM are also included for %comparison. 

%%%%%%%%%%%%%%%
\subsection{The basics of 2HDM}
\label{section-2HDM}

Here we shortly remind the basic aspects of the 2HDM that are relevant for the present computation.  
The 2HDM is the simplest extension of the SM  with two Higgs doublets,  $\Phi_1$ and $\Phi_2$,  instead of one doublet $\Phi$.  These two doublets are linear parametrizations of the four complex scalar fields (hence,  eight real scalar fields) defining the 2HDM  scalar sector.  They are usually defined as:
\begin{eqnarray}
\Phi_1 = \left( \begin{array}{c} \phi_1^+ \\ \frac{1}{\sqrt{2}} (v_1 +
    \rho_1 + i \eta_1) \end{array} \right) \;, \quad
\Phi_2 = \left( \begin{array}{c} \phi_2^+ \\ \frac{1}{\sqrt{2}} (v_2 +
    \rho_2 + i \eta_2) \end{array} \right) \;,
\label{eq:2hdmvevs}
\end{eqnarray}
where $v_1, v_2$ are the real vacuum expectation values  acquired by the fields
$\Phi_1, \Phi_2$, respectively, with $\tan\beta= v_2/v_1$ and they satisfy the 
relation $v = \sqrt{(v_1^2 +v_2^2)}$ where $v =246 \,  {\rm GeV}$ is again the SM vev.
The eight degrees of freedom above, $\phi_{1,2}^\pm$, $\rho_{1,2}$ and
$\eta_{1,2}$ are related to three Goldstone bosons, $G^\pm$ and $G^0$,
and five massive physical scalar fields: two $CP$-even scalar fields,
$h$ and $H$, one $CP$-odd one, $A$, and one charged pair, $H^\pm$.
Here the mixing angles $\alpha$ and $\beta$ diagonalize the $CP$-even and -odd
sectors,  respectively.  These rotations define the physical mass eigenstates,  $h$, $H$,  $A$ and $H^\pm$ in terms of the EW interaction eigenstates (or the other way around) and are given by\footnote{The relations among the two usual notations for the GBs are $G^{\pm}=- i \omega^{\pm}$,  $G^0=- \omega^0$}:
\begin{eqnarray}
\phi_1^\pm&=&\cos \beta \, G^\pm -\sin \beta \, H^\pm,\nonumber \\
\phi_2^\pm&=&\sin \beta \, G^\pm +\cos \beta \, H^\pm, \nonumber \\
\eta_1&=&\cos \beta \, G^0 -\sin \beta \, A, \nonumber \\
\eta_2&=&\sin \beta \, G^0 +\cos \beta \, A, \nonumber \\ 
\rho_1&=&\cos \alpha \, H -\sin \alpha \, h, \nonumber \\
\rho_2&=&\sin \alpha \, H +\cos \alpha \, h. 
\end{eqnarray}
The self-interactions among the above scalar fields are provided by the 2HDM potential which we assume here to have a $Z_2$ symmetry~\cite{Aoki:2009ha,Glashow:1976nt} meaning invariance under $\Phi_1\to\Phi_1$ and $\Phi_2\to-\Phi_2$.   This assumption ensures the absence of   flavor-changing neutral currents (FCNC) at the tree level, which are strongly discouraged by experimental measurements.   Furthermore,  we will allow 
this $Z_2$ symmetry to be only softly broken by the parameter $m_{12}^2$, which has dimensions of mass squared. 
The relevant 2HDM potential for the present work,  with the $Z_2$ soft-breaking included,  can be expressed in terms of the two doublets $\Phi_1$ and $\Phi_2$,  as follows~\cite{HHG,Aoki:2009ha,Branco:2011iw}:
\begin{equation}
\begin{split}
V_{\rm 2HDM}(\Phi_1,\Phi_2) = m_{11}^2 (\Phi_1^\dagger\Phi_1) + m_{22}^2 (\Phi_2^\dagger\Phi_2) - m_{12}^2 (\Phi_1^\dagger
\Phi_2 + \Phi_2^\dagger\Phi_1) + \frac{\lambda_1}{2} (\Phi_1^\dagger \Phi_1)^2 +
\frac{\lambda_2}{2} (\Phi_2^\dagger \Phi_2)^2 \\
 + \lambda_3
(\Phi_1^\dagger \Phi_1) (\Phi_2^\dagger \Phi_2) + \lambda_4
(\Phi_1^\dagger \Phi_2) (\Phi_2^\dagger \Phi_1) + \frac{\lambda_5}{2}
[(\Phi_1^\dagger \Phi_2)^2 +(\Phi_2^\dagger \Phi_1)^2] .
\end{split}
\label{2HDM_Higgs-pot}
\end{equation}
The full Lagrangian for the 2HDM (omitted here for shortness) is  invariant under $SU(2)_L \times U(1)_Y$ gauge transformations and it is usually presented in the generic $R_\xi$ covariant gauge.  The on-shell amplitudes computed within the 2HDM are gauge invariant as well.  For the computation here of the $WW \to hhh$ amplitude 
we will choose (as in the previous HEFT case) the unitary gauge,  where the GBs are absent and the EW gauge propagators are the proper ones for the unitary gauge.   

After the EW symmetry breaking,  $SU(2)_L \times U(1)_Y \to U(1)_{\rm em}$,  the minimization conditions for the above 2HDM potential lead to the existence of five physical Higgs bosons: two CP-even Higgs bosons $h$ and $H$,  one CP-odd Higgs boson $A$ and  two charged Higgs bosons $H^\pm$ with masses given by $m_h$,  $m_H$ (with $m_h < m_H$),  $m_A$ and $m_{H^\pm}$ respectively.
In addition,  the three would-be Goldstone bosons disappear from the physical spectrum and provide the needed physical masses for the EW gauge bosons,  $m_W$ and $m_Z$. 
We will also identify the $h$ state of the 2HDM with the Higgs boson discovered at the LHC with a mass $m_h= 125$ GeV~\cite{ATLAS:2012yve,CMS:2012qbp,ATLAS:2015yey}.  The other Higgs bosons will be assumed here to be heavier than the EW scale $v$,  a hypothesis which is well justified given the present tight experimental constraints~\cite{Arco:2020ucn,Arco:2022xum}. 

The previous potential also contains the self-interactions among the scalars of the 2HDM which are very relevant for the present work.  In addition,  the interactions of the Higgs bosons with the gauge bosons are given by the gauge invariant Lagrangian built with the covariant derivatives of the two doublets and the interactions with fermions are given by the Yukawa part of the 2HDM Lagrangian.  They are well known in the literature and  we do not explicit them here for shortness (see for instance~\cite{HHG,Aoki:2009ha,Branco:2011iw}). 
The set of Feynman rules within the 2HDM that are relevant for the present computation can be found,  for instance,  in \cite{ArcoGarcia:2023zjz}. 

In order to make predictions for observables, one may use different choices for the 2HDM input parameters.  In this work we choose the following 2HDM input parameters:
\begin{equation}
	v, \;\;\mh,\;\;\mHeavy,\;\;\mA,\;\;\mHp,\;\;\tanb,\;\;\cba,\;\;m_{12},
\label{eq-inputs}
\end{equation}
where we have adopted the shorthand notation $\cos x=c_x$, $\sin x=s_x$ and $\tan x=t_x$.  All the remaining 2HDM parameters and couplings are therefore derived quantities. This is a convenient choice since the on-shell scattering amplitude of our interest here $WW \to hhh$ will be written in terms of physical parameters like the on-shell masses for the external legs,  $m_W$ and $m_h$ and $v$ (or equivalently $G_F$) which will be assumed to be lighter than the heavy masses,  $m_H$,  $m_A$ and $m_{H^\pm}$. 

The choice of $\cba$ as an input parameter is motivated by the so-called \textit{alignment limit}, defined as ${\cba=0}$. 
Under this limit,  the interactions of $h$ with the gauge bosons and with the fermions recover their SM values.
For example,  the $h$ ($H$) coupling to $WW$ and $ZZ$ relative to the SM is given by $\sba$ ($\cba$).
Currently, the measurements of the Higgs boson signal strengths are compatible with the SM prediction (within the experimental uncertainties) and therefore, the parameter $\cba$ is constrained to be not far away from the alignment limit (see, for instance,\cite{Arco:2020ucn,  Arco:2022xum}).
We will assume here that this $\cba$ is in any case within the interval $|\cba|< 0.35$. 
This motivates considering two different scenarios for the 2HDM:  
1) 2HDM with alignment,  defined by setting ${\cba=0}$  and 2) quasi-alignment, defined by $\cba \neq 0$ but yet small  to be compatible with the experimental constraints.  As we will see in \secref{section-matching},  the solutions for the matching will differ in these two situations. 

It is convenient to recall  that taking the alignment condition does not imply the absence of BSM interactions, because several couplings, in addition to the SM ones,  remain non-vanishing when  $\cba=0$,  indeed,  some of them involving the light Higgs boson $h$.  For instance,  the couplings $hHH$,  $hAA$,  $hH^+H^-$, $ZHA$,  $\gamma H^+H^-$,  $hhHH$,  $hhAA$, 
$hhH^+H^-$,  and others do not vanish in the alignment limit.  This implies that the integration out of the heavy modes,  $H$, $A$ and $H^\pm$ may leave some non-decoupling effects that differentiate the 2HDM with respect to the SM via these couplings that could leave an important track at low energy observables.  This is our main motivation in this work and will be discussed in the following sections.    

One interesting phenomenological feature of the 2HDM compared to the SM is the existence of new triple and quartic interactions between the new scalar states.  
The derived values  of all these triple and quartic couplings should be written in terms of the input parameters chosen for the 2HDM,  given in \eqref{eq-inputs} in our case.
The tree-level Feynman rules of the scalar interactions involving the light Higgs that are relevant to this work  are summarized by:
\begin{eqnarray}
	i \Gamma_{hhh} &=& - 6 i \vev \lahhh, \nn \\
	%\label{eq-hhh} 
	%\\
	i \Gamma_{hhH} &=& - 2 i \vev \lahhH,\nn \\
	i \Gamma_{hH^+H^-} &=& -i \vev \lahHpHm, \nn \\
	i \Gamma_{hhhh} &=& - 6 i \lahhhh, \nn \\
	i\Gamma_{hhhH}&=&-6i\lambda_{hhhH}, \nn \\
i\Gamma_{hHH}&=&-2iv\lambda_{hHH},
\label{Gammas-2HDM}
\end{eqnarray}
where we are following the notation from~\cite{Arco:2020ucn,Arco:2022xum}.  Other scalar interactions involving the light Higgs like,  for instance, 
$\Gamma_{hAA}$,  do not participate in the present computation, and are not given explicitly here,  for shortness.  

One important aspect in this work is that these triple and quartic couplings above are not input parameters but instead they are derived parameters.  Thus,  once  the input parameters have been fixed to those in \eqref{eq-inputs},  the derived  couplings for the physical eigenstates, i.e.  the $\lambda_{\rm x}$ couplings,  are fixed in terms of the input parameters.  In particular,  at the tree level,  these derived $\lambda_{\rm x}$ couplings are given by:
%\begin{eqnarray}
 %\label{def-lahhh}
%	\vev^2\lambda_{hhh} &=& \sba\left(1+2\cba^2\right) \frac{\mh^2}{2} -\sba\cba^2\mbarc +\cba^3 \cot 2\beta \left(\mh^2-\mbarc\right) , \\
% \label{def-lahhH}
%	\vev^2\lambda_{hhH} &=& \frac{\cba}{2} \left( -2 \left(3 \cba^2-2\right) \mbarc-2 \cba \sba \cot 2\beta \left(-3 \mbarc+2 \mh^2+\mHeavy^2\right) %\right. \nn  \\
%	&&\biggl. +\left(2\cba^2-1\right) \left(2 \mh^2+\mHeavy^2\right) \biggr), \\
 %\label{def-lahHpHm}
%	\vev^2\lahHpHm &=& \left(\mh^2 +2 \mHp^2 -2 \mbarc \right)\sba + 2\cot2\beta \left( \mh^2 - \mbarc \right)\cba \,,  \\
 %\label{def-lahhhh} 
%	\vev^2\lahhhh &=& \frac{\mh^2}{2} +\frac{\cba^2}{2}\left(-4\sba^2\mbarc +4 \cba^2 \cot^2 2\beta \left(-\mbarc+\cba^2\mh^2+%\sba^2\mHeavy^2\right) \right.  \nn\\
%	&&\left.+4 \cba\sba \cot 2\beta \left(-2 \mbarc+\left(2\cba^2+1\right) \mh^2+\left(1-2 \cba^2\right) \mHeavy^2\right) \right.  \nn\\
%	&&\biggl.+4 \cba^4 \left(\mHeavy^2 - \mh^2\right) -4\cba^2\mHeavy^2 +3\mh^2 +\mHeavy^2\biggr) 
%\end{eqnarray}

%%%%%
\begin{align}
\label{lambdas}
v^2\lambda_{hhh} =&
s_{\beta-\alpha}\left(1+2c_{\beta-\alpha}^2\right)\frac{m_h^2}{2}
-s_{\beta-\alpha}c_{\beta-\alpha}^2\bar m^2
+c_{\beta-\alpha}^3\cot(2\beta)(m_h^2-\bar m^2),
\nn \\
v^2\lambda_{hhH} =&
\frac{c_{\beta-\alpha}}{2}
\Bigg(
-2(3c_{\beta-\alpha}^2-2)\bar m^2
-2c_{\beta-\alpha}s_{\beta-\alpha}\cot(2\beta)
\left(-3\bar m^2+2m_h^2+m_H^2\right)
\nn \\
&+\left(2c_{\beta-\alpha}^2-1\right)
\left(2m_h^2+m_H^2\right)
\Bigg),\notag  \nn \\
v^2\lambda_{hH^+H^-} =&
\left(m_h^2+2m_{H^\pm}^2-2\bar m^2\right)s_{\beta-\alpha}
+2\cot(2\beta)(m_h^2-\bar m^2)c_{\beta-\alpha},  \nn \\
v^2\lambda_{hhhh} =&
\frac{m_h^2}{2}
+\frac{c_{\beta-\alpha}^2}{2}
\Bigg(
-4s_{\beta-\alpha}^2\bar m^2
+4c_{\beta-\alpha}^2\cot^2(2\beta)
\left(
-\bar m^2+c_{\beta-\alpha}^2m_h^2+s_{\beta-\alpha}^2m_H^2
\right) \nn \\
&+4c_{\beta-\alpha}s_{\beta-\alpha}\cot(2\beta)
\left(
-2\bar m^2
+\left(2c_{\beta-\alpha}^2+1\right)m_h^2
+\left(1-2c_{\beta-\alpha}^2\right)m_H^2
\right)\notag   \nn \\
&+4c_{\beta-\alpha}^4(m_H^2-m_h^2)
-4c_{\beta-\alpha}^2m_H^2
+3m_h^2+m_H^2
\Bigg),\notag \nn \\
v^2\lambda_{hhhH} =&
-\frac{c_{\beta-\alpha}}{2}
\Bigg(
2s_{\beta-\alpha}^2c_{\beta-\alpha}^3\cot(2\beta)
\left(5m_h^2-2m_H^2\right)
-4s_{\beta-\alpha}^3c_{\beta-\alpha}^2
\left(-1+\cot^2(2\beta)\right)
(m_h^2-m_H^2)
 \nn \\
&
+2s_{\beta-\alpha}^4c_{\beta-\alpha}\cot(2\beta)
(m_h^2+2m_H^2)
-6s_{\beta-\alpha}^2c_{\beta-\alpha}\cot(2\beta)\bar m^2\notag
\nn \\
&
+s_{\beta-\alpha}^3
\left(m_h^2+m_H^2-2\bar m^2\right)
-2c_{\beta-\alpha}^3\cot(2\beta)
(m_h^2-\bar m^2)\notag
\nn \\
&
+s_{\beta-\alpha}c_{\beta-\alpha}^2
\Big(
\left(-3+4\cot^2(2\beta)\right)m_h^2
+m_H^2
+2\left(1-2\cot^2(2\beta)\right)\bar m^2
\Big)
\Bigg),\notag\\
v^2\lambda_{hHH} =&
\frac{s_{\beta-\alpha}}{2}
\Bigg(\left(m_h^2+2m_H^2-2\bar m^2\right)
s_{\beta-\alpha}^2
+2\left(m_h^2+2m_H^2-3\bar m^2\right)\cot(2\beta)
c_{\beta-\alpha}s_{\beta-\alpha}  \nn \\
&-\left(m_h^2+2m_H^2-4\bar m^2\right)c_{\beta -\alpha}^2
\Bigg),   
%\notag
\end{align}
where, to shorten the formulas,  we have used a new convenient parameter $\bar m$ defined as:
\begin{equation}
\bar m^2 \equiv \frac{m_{12}^2}{s_ \beta c_\beta} \,  .
\end{equation}
%%%%%%
Also,  to present  more compact formulas we have included some derived parameters in these formulas like $\sba$, $\sb$, $\cb$ and $\cot2\beta$, that are related with the input parameters, $\cba$ and $\tan \beta$,   by the following trigonometric identities:
\be
\sba=\sqrt{1-\cba^2}\,,\,\,
\sb=\frac{\tanb}{\sqrt{1+\tanb^2}}\,, \,\,
\cb=\frac{1}{\sqrt{1+\tanb^2}}\,,\,\,
\cot2\beta=\frac{1-\tanb^2}{2\tanb}\,.
\label{trigo}
\ee
For the discussion in the following sections it is interesting to display the specific values of the couplings above in the simplest scenario with alignment,  i.e.  for ${\cba=0}$,  which are named 
$\lambda_{\rm x}^{\rm al}$ here:
\begin{eqnarray}
	\vev^2\lambda_{hhh}^{\rm al} &=&  \frac{\mh^2}{2}\,, \nn \\
		\vev^2\lambda_{hhH}^{\rm al} &=&  0\,, \nn  \\
	\vev^2\lahHpHm^{\rm al}&=& \mh^2 +2 \mHp^2 -2 \mbarc\,, \nn   \\	
	\vev^2\lahhhh^{\rm al} &=& \frac{\mh^2}{2} ,\, \nn \\
	\vev^2 \lambda_{hhhH}^{\rm al} &=&  0\,, \nn  \\
	\vev^2 \lambda_{hHH}^{\rm al} &=& \frac{1}{2}(m_h^2 + 2 m_H^2 -2 \mbarc). 
\end{eqnarray}

We see  that in the alignment limit, $\lahhh$ and $\lahhhh$ tend to $\lambda=\mh^2/\left(2v^2\right)$,  which give respectively the triple and quartic Higgs couplings predicted by the SM.  The triple coupling of the two light Higgs bosons to one heavy Higgs boson vanishes in the alignment limit.  However, the coupling of one light Higgs boson to two charged Higgs bosons is not vanishing.   Regarding the above value of $\lahHpHm^{\rm al}$,  we confirm that the size of this triple coupling can be very large for large input values of $\mHp$.  For instance,  values of $\mHp \sim 800\,  {\rm GeV} $ can provide large couplings of $\lahHpHm^{\rm al}\sim {\cal O}(10)$ which, according to the  detailed analysis in~\cite{Arco:2020ucn,Arco:2022xum},  are still allowed by all the present theoretical and experimental constraints.  This is the situation we are interested in this work.  Then,  when doing a large mass expansion of the amplitudes in the following sections we simply mean an expansion in powers of a small dimensionless parameter $\sim (v/m_{\rm heavy})$ which should be convergent whenever  $v/m_{\rm heavy} \ll 1$.  Thus,  we have in mind that our forthcoming results for heavy boson masses,   $m_H$,  $m_A$ and $m_{H^\pm}$ which are collectively named $m_{\rm heavy}$,  should apply for these masses being above $v$ and close to the TeV scale.    

Within the 2HDM, the Higgs couplings to fermions also differ with respect to the SM. 
However, since in this work we only consider interactions in the bosonic sector of the 2HDM, they are not relevant to this work and we will not describe them here.

%%%%%%%%%%%%%%%%%%%%%%%%%%%%%%%
\section{Analytical results for $WW \to hhh$ amplitudes:  HEFT and 2HDM}
\label{section-amplitudes}
Here we present the analytical  results of  the scattering amplitudes for $WW\to hhh$ in both HEFT and 2HDM using the diagrammatic approach, i.e. identifying the contributions from each diagram in the unitary gauge. 
We consider the process $W^+(k_1)W^-(k_2)\to h(k_3)h(k_4)h(k_5)$, where $\epsilon_1$ and $\epsilon_2$ denote the polarization vectors of the incoming $W^+$ and $W^-$ bosons, respectively, and we use the following notation for the combinations of the external momenta:
\be
k_{ij}=k_i-k_j\,,\qquad q_{ij}=k_i+k_j\,.
\label{momentum}
\ee
The amplitudes of the diagrams that are related by permutations of the external momenta are written in terms of the amplitude of a representative diagram.  In this notation, $k_i\to k_j$ means that $k_i$ is replaced by $k_j$ in the expression of the representative amplitude, $k_i\leftrightarrow k_j$ means that these two momenta are exchanged, and when several replacements are indicated they are performed simultaneously. 

The following expressions in this section were obtained using FeynArts~\cite{FeynArts} and FormCalc~\cite{FormCalc-LT}.

%The relevant Feynman rules are collected in \tabref{relevant-FR} of \appref{App-Frules} (for a full set see [19–21]).

%%%%MJ
\subsection{$WW \to hhh$ within HEFT}
\begin{figure}[H]
	\centering
	\includegraphics[width=0.96\textwidth]{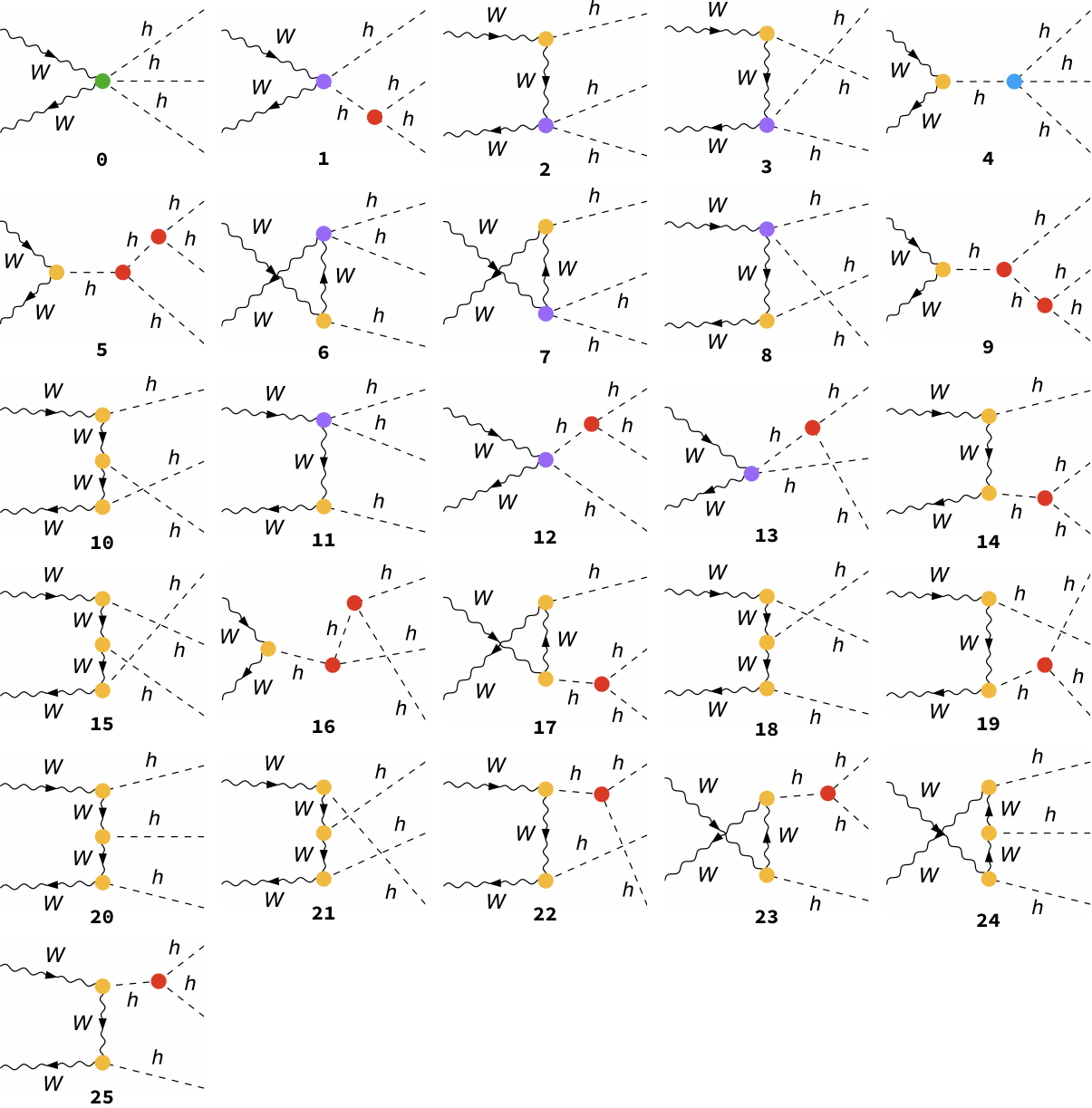}
	\caption{HEFT diagrams for $WW \to hhh$.}
	\label{HEFT-diagrams}
\end{figure}
The HEFT amplitude for $WW\to hhh$,  at the tree level in the unitary gauge,   receives contributions from 26 diagrams,  numbered here from 0 to 25,  which are shown in \figref{HEFT-diagrams}.   Notice that the different Higgs effective couplings are depicted by colored bullets: yellow for $WWh$ ($a$), purple for $WWhh$ ($b$), green for $WWhhh$ ($c$), red for $hhh$ ($\kappa_3$) and light blue for $hhhh$ ($\kappa_4$).  Notice also that the diagrams from 1 to 25 share  the same topologies as the corresponding ones in the SM.   In contrast,  the diagram 0 is not present in the SM case since the contact coupling of two W  with three Higgs bosons is nonexistent within the SM. 

We then find the total HEFT amplitude which is given by:
\bear
\amp^{\rm HEFT}&=&\amp^{\rm HEFT}_{0}+\amp^{\rm HEFT}_{1}+ \dots +\amp^{\rm HEFT}_{25},
\label{HEFT-total-amp}
\eear
where:
\bear
\amp^{\rm HEFT}_{0} &=& 6\,c\,\mw^2\vev^{-3}(\epsilon_1\cdot\epsilon_2)\,, \nn\\
\amp^{\rm HEFT}_{1} &=& \frac{6\,b\,\kappa_3\,\mh^2\mw^2\vev^{-3}}{q_{45}^2-\mh^2}(\epsilon_1\cdot\epsilon_2)\,, \nn\\
\amp^{\rm HEFT}_{2} &=& \frac{4\,a\,b\,\mw^4\vev^{-3}}{k_{13}^2-\mw^2}\Big((\epsilon_1\cdot\epsilon_2)-\mw^{-2}(\epsilon_1\cdot k_{13})(\epsilon_2\cdot k_{13})\Big)\,, \nn\\
\amp^{\rm HEFT}_{3} &=& \amp^{\rm HEFT}_{2}(k_3\to k_4)\,, \nn\\
\amp^{\rm HEFT}_{4} &=& \frac{6\,a\,\kappa_4\,\mh^2\mw^2\vev^{-3}}{q_{12}^2-\mh^2}(\epsilon_1\cdot\epsilon_2)\,, \nn\\
\amp^{\rm HEFT}_{5} &=& \frac{18\,a\,\kappa_3^2\,\mh^4\mw^2\vev^{-3}}{\left(q_{12}^2-\mh^2\right)\left(q_{34}^2-\mh^2\right)}(\epsilon_1\cdot\epsilon_2)\,, \nn\\
\amp^{\rm HEFT}_{6} &=& \amp^{\rm HEFT}_{2}(k_3\to k_5)\,, \nn\\
\amp^{\rm HEFT}_{7} &=& \amp^{\rm HEFT}_{2}(k_1\to k_2)\,, \nn\\
\amp^{\rm HEFT}_{8} &=& \amp^{\rm HEFT}_{2}(k_1\to k_2,\,k_3\to k_4)\,, \nn\\
\amp^{\rm HEFT}_{9} &=& \amp^{\rm HEFT}_{5}(k_3\to k_5)\,, \nn\\
\amp^{\rm HEFT}_{10} &=& \frac{8\,a^3\,\mw^6\vev^{-3}}{\left(k_{13}^2-\mw^2\right)\left(k_{24}^2-\mw^2\right)}\Big( (\epsilon_1\cdot\epsilon_2)-\mw^{-2}(\epsilon_1\cdot k_{24})(\epsilon_2\cdot k_{24})-\mw^{-2}(\epsilon_1\cdot k_{13})(\epsilon_2\cdot k_{13}) \nn\\
&&+\,\mw^{-4}(\epsilon_1\cdot k_{13})(\epsilon_2\cdot k_{24})(k_{24}\cdot k_{13})\Big)\,, \nn\\
\amp^{\rm HEFT}_{11} &=& \amp^{\rm HEFT}_{2}(k_1\to k_2,\,k_3\to k_5)\,, \nn\\
\amp^{\rm HEFT}_{12} &=& \amp^{\rm HEFT}_{1}(k_5\leftrightarrow k_3)\,, \nn\\
\amp^{\rm HEFT}_{13} &=& \amp^{\rm HEFT}_{1}(k_4\leftrightarrow k_3)\,, \nn\\
\amp^{\rm HEFT}_{14} &=& \frac{12\,a^2\,\kappa_3\,\mh^2\mw^4\vev^{-3}}{\left(k_{13}^2-\mw^2\right)\left(q_{45}^2-\mh^2\right)}\Big((\epsilon_1\cdot\epsilon_2)-\mw^{-2}(\epsilon_1\cdot k_{13})(\epsilon_2\cdot k_{13})\Big)\,, \nn\\
\amp^{\rm HEFT}_{15} &=& \amp^{\rm HEFT}_{10}(k_3\leftrightarrow k_4)\,, \nn\\
\amp^{\rm HEFT}_{16} &=& \amp^{\rm HEFT}_{5}(k_4\to k_5)\,, \nn\\
\amp^{\rm HEFT}_{17} &=& \amp^{\rm HEFT}_{14}(k_1\to k_2)\,, \nn\\
\amp^{\rm HEFT}_{18} &=& \amp^{\rm HEFT}_{10}(k_3\to k_4,\,k_4\to k_5)\,, \nn\\
\amp^{\rm HEFT}_{19} &=& \amp^{\rm HEFT}_{14}(k_3\leftrightarrow k_4)\,, \nn\\
\amp^{\rm HEFT}_{20} &=& \amp^{\rm HEFT}_{10}(k_4\to k_5)\,, \nn\\
\amp^{\rm HEFT}_{21} &=& \amp^{\rm HEFT}_{10}(k_3\to k_5)\,, \nn\\
\amp^{\rm HEFT}_{22} &=& \amp^{\rm HEFT}_{14}(k_1\to k_2,\,k_3\to k_4)\,, \nn\\
\amp^{\rm HEFT}_{23} &=& \amp^{\rm HEFT}_{14}(k_3\leftrightarrow k_5)\,, \nn\\
\amp^{\rm HEFT}_{24} &=& \amp^{\rm HEFT}_{10}(k_3\to k_5,\,k_4\to k_3)\,, \nn\\
\amp^{\rm HEFT}_{25} &=& \amp^{\rm HEFT}_{14}(k_1\to k_2,\,k_3\to k_5)\,.
\label{ampWWtohhh-HEFT}
\eear
First,  we have checked the agreement of these analytical results for the LO-HEFT amplitudes above with the LO part of the results in \cite{Domenech:2025gmn}. 
We have also checked that when setting in the formulas above the coefficient values to $a=b=\kappa_3=\kappa_4=1$ and $c=0$ one gets, as expected,  agreement with 
the analytical results of the SM amplitudes in the unitary gauge.  

For later convenience in this work,  we find it useful  to classify these 26 amplitudes above according to the different products of the HEFT couplings involving the $a$,  $b$, $c$, $\kappa_3$ and $\kappa_4$ parameters.  These HEFT amplitudes are placed within the following seven groups:
\bear
a^3 &\rightarrow& \{10,\,15,\,18,\,20,\,21,\,24\}\,, \nn\\
ab &\rightarrow& \{2,\,3,\,6,\,7,\,8,\,11\}\,,\nn\\
a^2\kappa_3 &\rightarrow& \{14,\,17,\,19,\,22,\,23,\,25\}\,,\nn\\
a\kappa_3^2 &\rightarrow& \{5,\,9,\,16\}\,,\nn\\
b\kappa_3 &\rightarrow& \{1,\,12,\,13\}\,,\nn\\
a\kappa_4 &\rightarrow& \{4\}\,,\nn\\
c &\rightarrow& \{0\}\,.
\label{HEFT-classification}
\eear
%%%MJ

\subsection{$WW \to hhh$ within 2HDM}
\begin{figure}[t]
	\centering
	\includegraphics[width=0.9\textwidth]{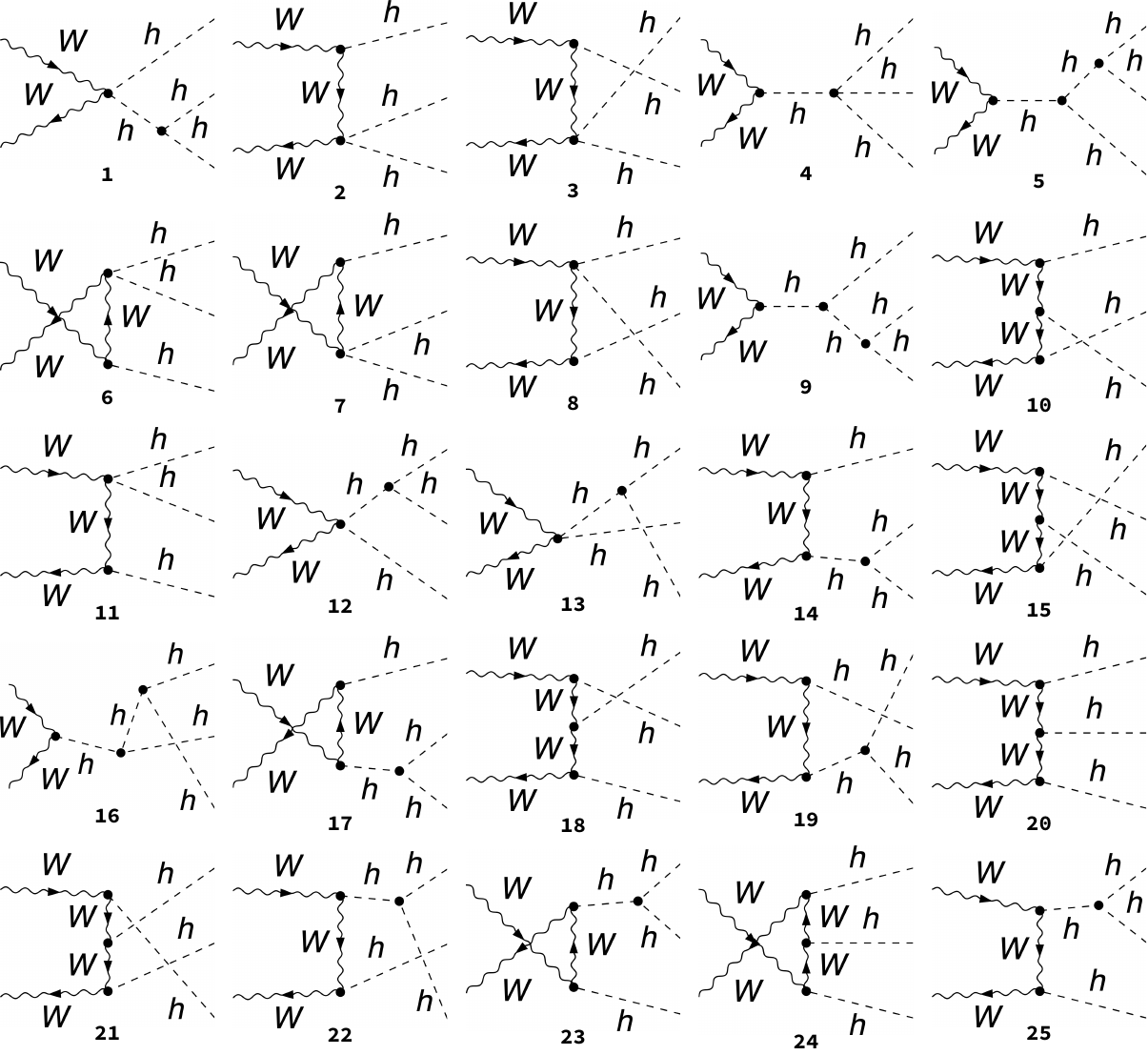}
	\caption{The 25 2HDM diagrams with the same topologies as the HEFT diagrams, involving only the light Higgs boson $h$.}
	\label{2HDM-diagrams-1-25}
\end{figure}
\begin{figure}[t]
	\centering
	\includegraphics[width=0.9\textwidth]{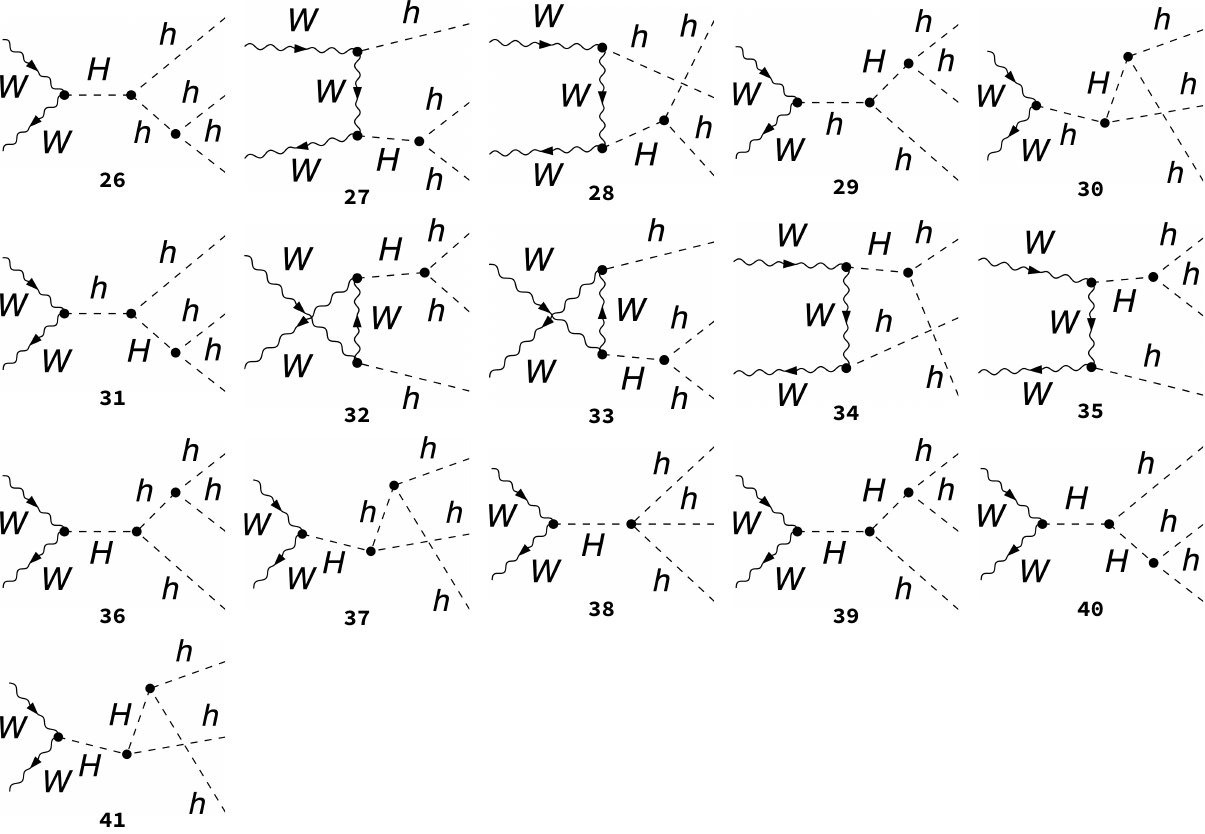}
	\caption{The 16 additional 2HDM diagrams involving couplings to the heavy Higgs boson $H$.}
	\label{2HDM-diagrams-26-41}
\end{figure}
\begin{figure}[t]
	\centering
	\includegraphics[width=0.9\textwidth]{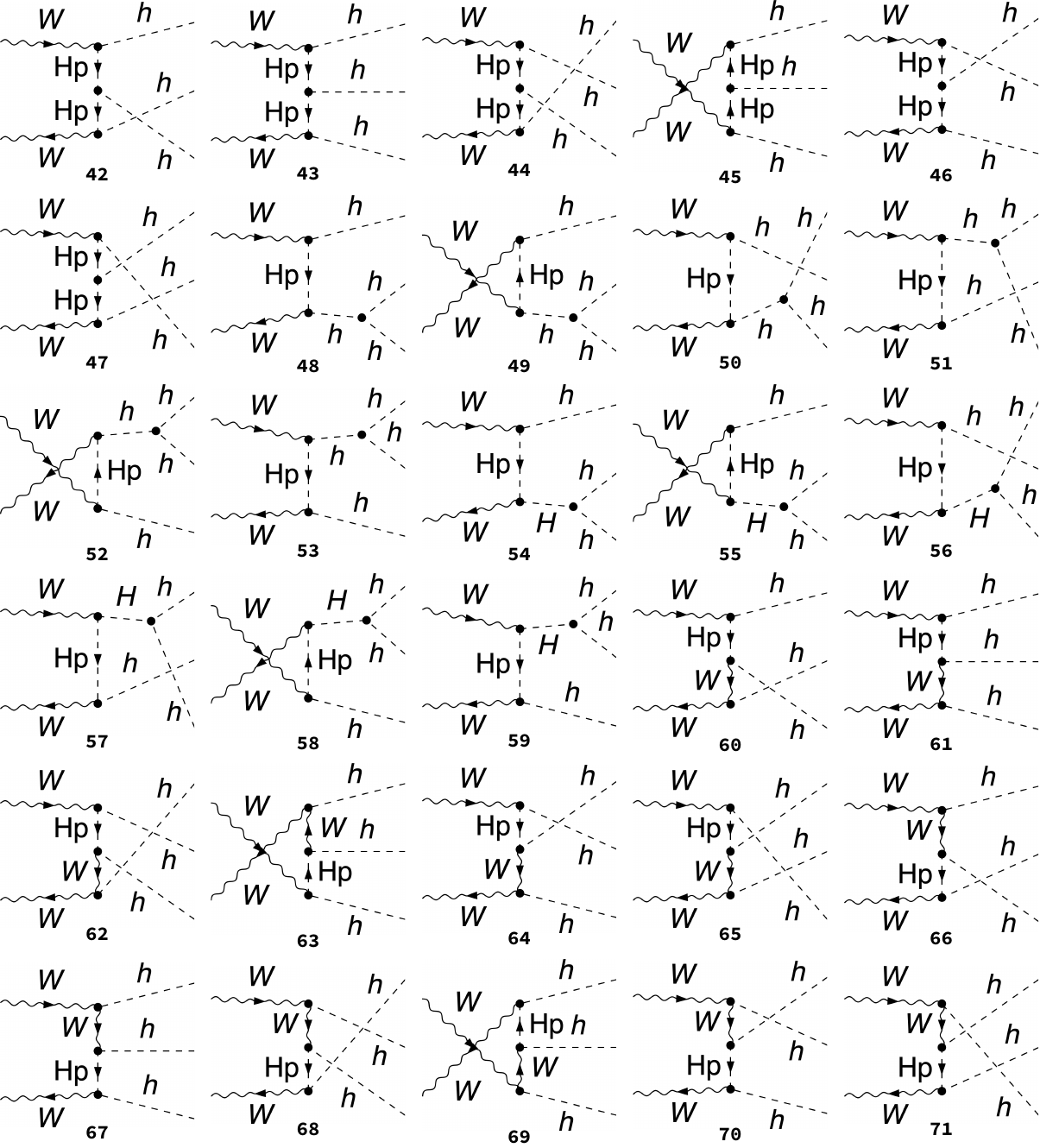}
	\caption{The remaining 30 2HDM diagrams, involving couplings to the heavy neutral Higgs boson $H$ and the charged Higgs bosons $H^\pm$ (these are labeled Hp in the figure). }
	\label{2HDM-diagrams-42-71}
\end{figure}

Within the 2HDM,  in the unitary gauge,  there are 71 diagrams contributing at the tree level to the scattering process $WW\to hhh$.  These diagrams are shown in  \figref{2HDM-diagrams-1-25},   \figref{2HDM-diagrams-26-41} and \figref{2HDM-diagrams-42-71}. 
In order to facilitate the procedure of solving the matching equations in the next section among the 2HDM and the HEFT amplitudes,  and to identify more easily the non-decoupling contributions in the heavy-mass limit that we are interested in,  we find it convenient to classify these 2HDM diagrams into the following three groups. 
The first group contains 25 diagrams, shown in \figref{2HDM-diagrams-1-25}, which share the same topologies as those in the HEFT (and in the SM) and involve only the light Higgs boson $h$, namely,  no heavy Higgs bosons are propagating in these diagrams.  In the alignment limit,  the Higgs self-couplings $hhh$ and $hhhh$ recover their SM values and therefore these diagrams converge to the SM ones.  Note again that the HEFT diagram 0, corresponding to the $WWhhh$ contact interaction, has no counterpart in the 2HDM nor in the SM.

The second group consists of 16 diagrams, shown in \figref{2HDM-diagrams-26-41}, which involve the heavy neutral CP-even Higgs boson $H$ through interaction vertices of the types $hhH$, $HHh$, and $Hhhh$.  We will see later that these diagrams are the only ones that contribute to the final non-decoupling effects in the heavy-mass limit.

Finally, the third group includes 30 diagrams, shown in \figref{2HDM-diagrams-42-71},  which contain heavy Higgs bosons propagating,  neutral $H$ and charged $H^\pm$,   and with at most one scalar coupling $\lambda_x$.  As we will see later, these diagrams do not contribute to the non-decoupling effects we are looking for.

%\newpage
The amplitudes of the 71 diagrams are given by:
\bear
\amp^{\rm 2HDM}_{1} &=& \frac{12\,\vev^{-1}\lahhh\,\mw^2}{q_{45}^2-\mh^2}(\epsilon_1\cdot\epsilon_2)\,, \nn\\
\amp^{\rm 2HDM}_{2} &=& \frac{4\,\mw^4\sba\vev^{-3}}{k_{13}^2-\mw^2}\Big((\epsilon_1\cdot\epsilon_2)-\mw^{-2}(\epsilon_1\cdot k_{13})(\epsilon_2\cdot k_{13})\Big)\,, \nn\\
\amp^{\rm 2HDM}_{3} &=& \amp^{\rm 2HDM}_{2}(k_3\to k_4)\,, \nn\\
\amp^{\rm 2HDM}_{4} &=& \frac{12\,\lahhhh\,\mw^2\vev^{-1}\sba}{q_{12}^2-\mh^2}(\epsilon_1\cdot\epsilon_2)\,, \nn\\
\amp^{\rm 2HDM}_{5} &=& \frac{72\,\lahhh^2\,\mw^2\vev\,\sba}{\left(q_{12}^2-\mh^2\right)\left(q_{34}^2-\mh^2\right)}(\epsilon_1\cdot\epsilon_2)\,, \nn\\
\amp^{\rm 2HDM}_{6} &=& \amp^{\rm 2HDM}_{2}(k_3\to k_5)\,, \nn\\
\amp^{\rm 2HDM}_{7} &=& \amp^{\rm 2HDM}_{2}(k_1\to k_2)\,, \nn\\
\amp^{\rm 2HDM}_{8} &=& \amp^{\rm 2HDM}_{2}(k_1\to k_2,\,k_3\to k_4)\,, \nn\\
\amp^{\rm 2HDM}_{9} &=& \amp^{\rm 2HDM}_{5}(k_3\to k_5)\,, \nn\\
\amp^{\rm 2HDM}_{10} &=& \frac{8\,\sba^3\,\mw^6\vev^{-3}}{\left(k_{13}^2-\mw^2\right)\left(k_{24}^2-\mw^2\right)}\Big( (\epsilon_1\cdot\epsilon_2)-\mw^{-2}(\epsilon_1\cdot k_{24})(\epsilon_2\cdot k_{24})-\mw^{-2}(\epsilon_1\cdot k_{13})(\epsilon_2\cdot k_{13}) \nn\\
&&+\,\mw^{-4}(\epsilon_1\cdot k_{13})(\epsilon_2\cdot k_{24})(k_{24}\cdot k_{13})\Big)\,, \nn\\
\amp^{\rm 2HDM}_{11} &=& \amp^{\rm 2HDM}_{2}(k_1\to k_2,\,k_3\to k_5)\,, \nn\\
\amp^{\rm 2HDM}_{12} &=& \amp^{\rm 2HDM}_{1}(k_5\to k_3)\,, \nn\\
\amp^{\rm 2HDM}_{13} &=& \amp^{\rm 2HDM}_{1}(k_4\to k_3)\,, \nn\\
\amp^{\rm 2HDM}_{14} &=& \frac{24\,\lahhh\,\mw^4\vev^{-1}\sba^2}{\left(k_{13}^2-\mw^2\right)\left(q_{45}^2-\mh^2\right)}\Big((\epsilon_1\cdot\epsilon_2)-\mw^{-2}(\epsilon_1\cdot k_{13})(\epsilon_2\cdot k_{13})\Big)\,, \nn\\
\amp^{\rm 2HDM}_{15} &=& \amp^{\rm 2HDM}_{10}(k_3\leftrightarrow k_4)\,, \nn\\
\amp^{\rm 2HDM}_{16} &=& \amp^{\rm 2HDM}_{5}(k_4\to k_5)\,, \nn\\
\amp^{\rm 2HDM}_{17} &=& \amp^{\rm 2HDM}_{14}(k_1\to k_2)\,, \nn\\
\amp^{\rm 2HDM}_{18} &=& \amp^{\rm 2HDM}_{10}(k_3\to k_4,\,k_4\to k_5)\,, \nn\\
\amp^{\rm 2HDM}_{19} &=& \amp^{\rm 2HDM}_{14}(k_3\leftrightarrow k_4)\,, \nn\\
\amp^{\rm 2HDM}_{20} &=& \amp^{\rm 2HDM}_{10}(k_4\to k_5)\,, \nn\\
\amp^{\rm 2HDM}_{21} &=& \amp^{\rm 2HDM}_{10}(k_3\to k_5)\,, \nn\\
\amp^{\rm 2HDM}_{22} &=& \amp^{\rm 2HDM}_{14}(k_1\to k_2,\,k_3\to k_4)\,, \nn\\
\amp^{\rm 2HDM}_{23} &=& \amp^{\rm 2HDM}_{14}(k_3\leftrightarrow k_5)\,, \nn\\
\amp^{\rm 2HDM}_{24} &=& \amp^{\rm 2HDM}_{10}(k_3\to k_5,\,k_4\to k_3)\,, \nn\\
\amp^{\rm 2HDM}_{25} &=& \amp^{\rm 2HDM}_{14}(k_1\to k_2,\,k_3\to k_5)\,, \nn\\
\amp^{\rm 2HDM}_{26} &=& \frac{24\,\lahhh\,\lahhH\,\mw^2\vev\,\cba}{\left(q_{45}^2-\mh^2\right)\left(q_{12}^2-\mHeavy^2\right)}(\epsilon_1\cdot\epsilon_2)\,, \nn\\
\amp^{\rm 2HDM}_{27} &=& \frac{8\,\lahhH\,\mw^4\vev^{-1}\sba\,\cba}{\left(k_{13}^2-\mw^2\right)\left(q_{45}^2-\mHeavy^2\right)}\Big((\epsilon_1\cdot\epsilon_2)-\mw^{-2}(\epsilon_1\cdot k_{13})(\epsilon_2\cdot k_{13})\Big)\,, \nn\\
\amp^{\rm 2HDM}_{28} &=& \amp^{\rm 2HDM}_{27}(k_3\to k_4)\,, \nn
\eear
\bear
\amp^{\rm 2HDM}_{29} &=& \frac{8\,\lahhH^2\,\mw^2\vev\,\sba}{\left(q_{12}^2-\mh^2\right)\left(q_{34}^2-\mHeavy^2\right)}(\epsilon_1\cdot\epsilon_2)\,, \nn\\
\amp^{\rm 2HDM}_{30} &=& \amp^{\rm 2HDM}_{29}(k_4\to k_5)\,, \nn\\
\amp^{\rm 2HDM}_{31} &=& \amp^{\rm 2HDM}_{29}(k_3\to k_5)\,, \nn\\
\amp^{\rm 2HDM}_{32} &=& \amp^{\rm 2HDM}_{27}(k_3\to k_5)\,, \nn\\
\amp^{\rm 2HDM}_{33} &=& \amp^{\rm 2HDM}_{27}(k_1\to k_2)\,, \nn\\
\amp^{\rm 2HDM}_{34} &=& \amp^{\rm 2HDM}_{27}(k_1\to k_2,\,k_3\to k_4)\,, \nn\\
\amp^{\rm 2HDM}_{35} &=& \amp^{\rm 2HDM}_{27}(k_1\to k_2,\,k_3\to k_5)\,, \nn\\
\amp^{\rm 2HDM}_{36} &=& \amp^{\rm 2HDM}_{26}(k_5\to k_3)\,, \nn\\
\amp^{\rm 2HDM}_{37} &=& \amp^{\rm 2HDM}_{26}(k_4\to k_3)\,, \nn\\
\amp^{\rm 2HDM}_{38} &=& \frac{12\,\lambda_{hhhH}\,\mw^2\vev^{-1}\cba}{q_{12}^2-\mHeavy^2}(\epsilon_1\cdot\epsilon_2)\,, \nn\\
\amp^{\rm 2HDM}_{39} &=& \frac{8\,\lahhH\,\lambda_{HHh}\,\mw^2\vev\,\cba}{\left(q_{12}^2-\mHeavy^2\right)\left(q_{34}^2-\mHeavy^2\right)}(\epsilon_1\cdot\epsilon_2)\,, \nn\\
\amp^{\rm 2HDM}_{40} &=& \amp^{\rm 2HDM}_{39}(k_3\to k_5)\,, \nn\\
\amp^{\rm 2HDM}_{41} &=& \amp^{\rm 2HDM}_{39}(k_4\to k_5)\,, \nn\\
\amp^{\rm 2HDM}_{42} &=& \frac{4\,\cba^2\,\mw^2\vev^{-1}\lahHpHm}{\left(k_{13}^2-\mHp^2\right)\left(k_{24}^2-\mHp^2\right)}(\epsilon_1\cdot k_{13})(\epsilon_2\cdot k_{24})\,, \nn\\
\amp^{\rm 2HDM}_{43} &=& \amp^{\rm 2HDM}_{42}(k_4\to k_5)\,, \nn\\
\amp^{\rm 2HDM}_{44} &=& \amp^{\rm 2HDM}_{42}(k_3\leftrightarrow k_4)\,, \nn\\
\amp^{\rm 2HDM}_{45} &=& \amp^{\rm 2HDM}_{42}(k_3\to k_5,\,k_4\to k_3)\,, \nn\\
\amp^{\rm 2HDM}_{46} &=& \amp^{\rm 2HDM}_{42}(k_3\to k_4,\,k_4\to k_5)\,, \nn\\
\amp^{\rm 2HDM}_{47} &=& \amp^{\rm 2HDM}_{42}(k_3\to k_5)\,, \nn\\
\amp^{\rm 2HDM}_{48} &=& -\frac{24\,\cba^2\,\mw^2\vev^{-1}\lahhh}{\left(k_{13}^2-\mHp^2\right)\left(q_{45}^2-\mh^2\right)}(\epsilon_1\cdot k_{13})(\epsilon_2\cdot k_{13})\,, \nn\\
\amp^{\rm 2HDM}_{49} &=& \amp^{\rm 2HDM}_{48}(k_1\to k_2)\,, \nn\\
\amp^{\rm 2HDM}_{50} &=& \amp^{\rm 2HDM}_{48}(k_3\leftrightarrow k_4)\,, \nn\\
\amp^{\rm 2HDM}_{51} &=& \amp^{\rm 2HDM}_{48}(k_1\to k_2,\,k_3\leftrightarrow k_4)\,, \nn\\
\amp^{\rm 2HDM}_{52} &=& \amp^{\rm 2HDM}_{48}(k_3\leftrightarrow k_5)\,, \nn\\
\amp^{\rm 2HDM}_{53} &=& \amp^{\rm 2HDM}_{48}(k_1\to k_2,\,k_3\leftrightarrow k_5)\,, \nn\\
\amp^{\rm 2HDM}_{54} &=& -\frac{8\,\lahhH\,\mw^2\vev^{-1}\cba\,\sba}{\left(k_{13}^2-\mHp^2\right)\left(q_{45}^2-\mHeavy^2\right)}(\epsilon_1\cdot k_{13})(\epsilon_2\cdot k_{13})\,, \nn\\
\amp^{\rm 2HDM}_{55} &=& \amp^{\rm 2HDM}_{54}(k_1\to k_2)\,, \nn\\
\amp^{\rm 2HDM}_{56} &=& \amp^{\rm 2HDM}_{54}(k_3\leftrightarrow k_4)\,, \nn\\
\amp^{\rm 2HDM}_{57} &=& \amp^{\rm 2HDM}_{54}(k_1\to k_2,\,k_3\leftrightarrow k_4)\,, \nn\\
\amp^{\rm 2HDM}_{58} &=& \amp^{\rm 2HDM}_{54}(k_3\leftrightarrow k_5)\,, \nn\\
\amp^{\rm 2HDM}_{59} &=& \amp^{\rm 2HDM}_{54}(k_1\to k_2,\,k_3\leftrightarrow k_5)\,, \nn\\
\amp^{\rm 2HDM}_{60} &=& \frac{8\,\mw^4\sba\,\cba^2\vev^{-3}}{\left(k_{13}^2-\mHp^2\right)\left(k_{24}^2-\mw^2\right)}(\epsilon_1\cdot k_{13})\left((\epsilon_2\cdot k_{24})(k_1\cdot k_3)\mw^{-2}+(\epsilon_2\cdot k_{25})\right)\,, \nn\\
\amp^{\rm 2HDM}_{61} &=& \amp^{\rm 2HDM}_{60}(k_4\leftrightarrow k_5)\,, \nn\\
\amp^{\rm 2HDM}_{62} &=& \amp^{\rm 2HDM}_{60}(k_3\leftrightarrow k_4)\,, \nn\\
\amp^{\rm 2HDM}_{63} &=& \amp^{\rm 2HDM}_{60}(k_3\to k_5,\,k_4\to k_3,\,k_5\to k_4)\,, \nn\\
\amp^{\rm 2HDM}_{64} &=& \amp^{\rm 2HDM}_{60}(k_3\to k_4,\,k_4\to k_5,\,k_5\to k_3)\,, \nn\\
\amp^{\rm 2HDM}_{65} &=& \amp^{\rm 2HDM}_{60}(k_3\leftrightarrow k_5)\,, \nn
\eear
\bear
\amp^{\rm 2HDM}_{66} &=& \frac{8\,\mw^4\sba\,\cba^2\vev^{-3}}{\left(k_{24}^2-\mHp^2\right)\left(k_{13}^2-\mw^2\right)}(\epsilon_2\cdot k_{24})\left((\epsilon_1\cdot k_{13})(k_2\cdot k_4)\mw^{-2}+(\epsilon_1\cdot k_{15})\right)\,, \nn\\
\amp^{\rm 2HDM}_{67} &=& \amp^{\rm 2HDM}_{66}(k_4\leftrightarrow k_5)\,, \nn\\
\amp^{\rm 2HDM}_{68} &=& \amp^{\rm 2HDM}_{66}(k_3\leftrightarrow k_4)\,, \nn\\
\amp^{\rm 2HDM}_{69} &=& \amp^{\rm 2HDM}_{66}(k_3\to k_5,\,k_4\to k_3,\,k_5\to k_4)\,, \nn\\
\amp^{\rm 2HDM}_{70} &=& \amp^{\rm 2HDM}_{66}(k_3\to k_4,\,k_4\to k_5,\,k_5\to k_3)\,, \nn\\
\amp^{\rm 2HDM}_{71} &=& \amp^{\rm 2HDM}_{66}(k_3\leftrightarrow k_5)\,.
\label{ampWWtohhh-2HDM}
\eear
%%%%%%%%%%%%%%%%%%%%%%%%%%%%%%%
\section{Matching HEFT and 2HDM amplitudes}
\label{section-matching}
In this section we perform the matching between the HEFT and the 2HDM amplitudes for the process $WW\to hhh$ and we will do it at the tree level. 
Our goal is to determine the LO-HEFT coefficients $a$, $b$, $\kappa_3$, $\kappa_4$ and $c$ in terms of the input parameters of the 2HDM.  Concretely  we will provide the solutions to the matching in terms of the values of $\Delta a$, $\Delta b$, $\Delta\kappa_3$, $\Delta\kappa_4$ and $c$ (recall that $c=0$ in the SM),  as functions of the 2HDM input parameters of \eqref{eq-inputs}, namely,  $\vev$, $\mh$, $\cba$, $\tanb$ and $m_{12}$.  The values of these coefficients will summarize the non-decoupling effects that we are looking for,  namely,  those which are left from the heavy modes at low energies where the Effective Field Theory operates.   By non-decoupling effects here we mean  those that do not go as inverse powers of the heavy masses and therefore do not vanish in the $m_{\rm heavy} \to \infty$ limit.   

Since the HEFT contains as explicit degrees of freedom only the light modes,  i.e.  only the light Higgs boson $h$ in the Higgs sector,  it should reproduce the predictions of the 2HDM for $WW \to hhh$ once the heavy states $H$, $A$ and $H^\pm$ have been integrated out.  This behaviour is guaranteed if we require both theories to give the same amplitudes for $WW\to hhh$ when the masses of the heavy states, which we denote collectively by $m_{\rm heavy}$, are  larger than any other EW scale involved  in the process, i.e.,
\be
m_{\rm heavy}\gg m_{\rm EW}\,,
\ee
where $m_{\rm EW}$ stands generically for $\mw$, $\mz$, $\mh$, $\vev$, $m_{12}$ and also for the energy of the scattering, $\sqrt{s}$.
The matching condition is then written as:
\be
\amp^{\rm HEFT}=\amp^{\rm 2HDM}_{\rm heavy} \,,
\label{eq-ampmatching}
\ee
where by $\amp^{\rm 2HDM}_{\rm heavy}$ we mean  the 2HDM amplitude once it is expanded in powers of the small dimensionless parameter $m_{\rm EW}^2/m_{\rm heavy}^2$. 
Since we are interested in the non-decoupling effects of the heavy Higgs bosons, we only keep the terms of this expansion that do not vanish when $m_{\rm heavy}\to\infty$,  which as we will demonstrate next,  are those of order $(m_{\rm EW}^2/m_{\rm heavy}^2)^0$.   Notice also that, as we will check explicitly,  the total amplitude does not contain positive powers of $(m_{\rm heavy}^2)/(m_{\rm EW}^2)$,  as these potential contributions are canceled between diagrams,  and therefore the expansion is convergent and well defined. 
The next terms in the expansion of order $(m_{\rm EW}^2/m_{\rm heavy}^2)^n$ with $n=1,2,\ldots$ share a decoupling behaviour (i.e.  they vanish in the $m_{\rm heavy}\to\infty$ limit) which are not of our interest here since they are subleading and therefore will not be computed here. 

In order to make the matching process as simple as possible,  we organize it following the classification of the HEFT diagrams in \eqref{HEFT-classification}, and we solve it sequentially, from the groups of diagrams that depend on a single HEFT coefficient to those that depend on several of them.  For each group we compare the amplitudes with the same Lorentz structure in the 2HDM, which allows us to isolate the coupling structures associated to the HEFT coefficients.
We first obtain $a$ from the diagrams proportional to $a^3$. Then we use the diagrams proportional to $ab$ to get $b$. 
The coefficient $\kappa_3$ appears in three groups of diagrams, proportional to $b\kappa_3$, $a\kappa_3^2$ and $a^2\kappa_3$. We solve for $\kappa_3$ with the first one, and the other two serve as independent checks. 
Finally, $\kappa_4$ and $c$ have to be obtained at the same time, since the 2HDM diagrams that generate the $\kappa_4$ structure also produce, in the heavy mass limit, a contact term. 

\subsection{Large $m_{\rm heavy}$ expansion  of the 2HDM amplitudes and solutions to the matching}
Here we first analize the large $m_{\rm heavy}$ expansion  of the 2HDM amplitude for the $WW \to hhh$ process  and then we solve the matching equation \eqref{eq-ampmatching}. 

This large $m_{\rm heavy}$ expansion is performed for each of the 71 diagrams contributing to $\amp^{\rm 2HDM}$.  Since  the dependence on $m_{\rm heavy}$ only appears via the heavy propagators and the values of $\lambda_x$ in the numerators of these amplitudes,  our systematic method for this expansion is: 1)  to expand each heavy propagator in powers of $p^2/m_{\rm heavy}^2$,  where $p^2$ is the corresponding momentum squared in the propagator (i.e.  $k_{ij}^2$ or $q_{ij}^2$,  in our notation defined in \eqref{momentum}), and 2) follow the specific dependence on $m_{\rm heavy}^2$ in the numerator from the product of the various $\lambda_x$'s involved. 

Specifically,  we take first into account the typical expansion of each heavy propagator: 
\be
\frac{1}{p^2-m_{\rm heavy}^2}
=
-\frac{1}{m_{\rm heavy}^2}
\left[
1+\frac{p^2}{m_{\rm heavy}^2}
+\mathcal{O}\left(
\left(\frac{p^2}{m_{\rm heavy}^2}\right)^2
\right)
\right],
\ee
and,  second,  we follow the track of the $\lambda_x$'s which provide the ${\cal O}(m_{\rm heavy}^2)/m_{EW}^2$ contributions in the numerator,  concretely,  $\lambda_{hhH}$,  $\lambda_{hHH}$,  $\lambda_{hH^+H^-}$, $\lambda_{hhhH}$ and $\lambda_{hhhh}$.  The remaining $\lambda_{hhh}$ does not depend on $m_{\rm heavy}$.  

From this detailed analysis of the $m_{\rm heavy}^2$ dependence we conclude the following features which we analize in sets of diagrams as given in figures  \figref{2HDM-diagrams-1-25},   \figref{2HDM-diagrams-26-41} and \figref{2HDM-diagrams-42-71}. 

First set: Since diagrams 1 to 25 in the first set in  \figref{2HDM-diagrams-1-25} do not contain heavy propagators,  the only dependence on $m_{\rm heavy}^2$ comes from diagram 4 via the $\lambda_{hhhh}$ coupling.  This diagram therefore provides a potential non-decoupling effect with quadratic dependence on $m_{\rm heavy}$.   However,  this contribution is not physical since it will not appear in the total amplitude when adding all diagrams together.  Concretely,  we have checked that this ${\cal O}(m_{\rm heavy}^2)/m_{EW}^2$  contribution from diagram 4 cancels with the 
${\cal O}(m_{\rm heavy}^2)/m_{EW}^2$  contribution from diagrams 29, 30 and 31 of the second set in \figref{2HDM-diagrams-26-41}.  Thus, this first set does not contribute at the end to the final non-decoupling effects we are looking for.

Third set: The diagrams 42 to 71 in the third set in \figref{2HDM-diagrams-42-71} all contain heavy propagators.   The diagrams from 42 to 47  contain two heavy propagators (of charged Higgs bosons) and one $\lambda_x$,  therefore they  vanish in  the $m_{\rm heavy} \to \infty$ limit. 
Diagrams from 48 to 53,  and from 60 to 71  contain one heavy propagator (of charged Higgs bosons) and do not involve any of the $\lambda_x$'s carrying the $m_{\rm heavy}^2$ dependence.  Thus,  they all vanish as well in the 
$m_{\rm heavy} \to \infty$ limit.  Finally,  the diagrams from 54 to 59  contain two heavy propagators (one charged Higgs and one neutral Higgs) and involve just one  $\lambda_x$ with $m_{\rm heavy}^2$ dependence,  then they also vanish in the $m_{\rm heavy} \to \infty$ limit.  Therefore,  this third set  does not contribute either to the non-decoupling effects.

Second set: The diagrams from 26 to 41 in the second set in  \figref{2HDM-diagrams-26-41} are the relevant ones for our objective.  The diagrams  26, 27,  28,  32,  33,  34,  35,  36,  37 and 38 contain one heavy propagator (of neutral Higgs)  and one $\lambda_x$ with  $m_{\rm heavy}^2$ dependence,  thus they provide  $(m_{\rm EW}^2/m_{\rm heavy}^2)^0$ contributions in the expansion.  The diagrams 39, 40 and 41 contain two heavy propagators (of neutral Higgs) and two $\lambda_x$ with  $m_{\rm heavy}^2$ dependence each,  thus they also provide $(m_{\rm EW}^2/m_{\rm heavy}^2)^0$ contributions in the expansion.  Finally,  diagrams 29,  30 and 31 contain one heavy propagator (of neutral Higgs) and two $\lambda_x$ with  $m_{\rm heavy}^2$ dependence,  therefore,  they provide $(m_{\rm EW}^2/m_{\rm heavy}^2)^0$ contributions and also  ${\cal O}(m_{\rm heavy}^2/m_{\rm EW}^2)$ terms.  All diagrams in this set then contribute to the non-decoupling effects we are interested in.  

%For diagrams 26--28 and 32--41, it suffices to retain only the leading term in the propagator expansion. Higher-order terms are suppressed by additional powers of $1/m_{\rm heavy}^2$ and vanish in the heavy-mass limit. At this leading order, the suppression from the heavy $H$ propagators is exactly compensated by the $\mHeavy^2$ growth of the corresponding couplings, leaving finite non-decoupling contributions.

%By contrast, for diagrams 29--31, the expansion must be carried out up to $\mathcal{O}(k/m_{\rm heavy}^2)$. This sub-leading term yields an additional contribution to the contact interaction that would be completely missed if truncated at leading order. Furthermore, these diagrams generate terms proportional to $\mHeavy^2$ that exactly cancel the corresponding heavy-mass contributions from diagram 4 when the total amplitude is assembled. Finally, diagrams 42--71 do not contribute to the non-decoupling effects, as the mass dependence of their numerators does not compensate for the suppression introduced by the heavy
%propagators.

We report next the results of the large $m_{\rm heavy}$ expansion for the diagrams in the second set which are the only relevant ones for our purpose here. The amplitudes for diagrams in this set from 26 to 41 in the heavy-mass limit are shown in \eqref{ampWWtohhh-2HDM-heavy}.  As can be checked easily in this result,  the contributions of ${\cal O}(m_H^2)$ from diagrams 29, 30 and 31 in set two cancel the ones from diag 4 in set one, as announced.  
Notice also that,  as expected for the non-decoupling contributions,  all these amplitudes are non-vanishing in the $m_{\rm heavy} \to \infty$ limit since they do not decrease with inverse powers of the heavy mass. 

%Let us first describe how the different sets of 2HDM diagrams behave in the heavy mass limit.
%The diagrams 1-25 do not contain heavy propagators, and their amplitudes stay as they are in the previous section. 
%The only exception is diagram 4, which contains the quartic coupling $\lahhhh$ and therefore terms proportional to $\mHeavy^2$, see \eqref{lambdas}. 
%In the diagrams 26-41 the suppression from the (one or two) propagators of $H$ is compensated by the growth with $m_{\rm heavy}$ of the couplings $\lahhH$, $\lambda_{HHh}$ and $\lambda_{hhhH}$, so that these diagrams in principle give finite contributions of ${\cal O}(m_{\rm heavy}^0)$. 
%Moreover, the diagrams 29, 30 and 31 also contain terms proportional to $\mHeavy^2$, which cancel exactly against the one from diagram 4 when they are added. 
%Finally, the diagrams 42-71, with charged Higgs bosons, do not contribute to the matching at the order we consider. 

%The amplitudes of the diagrams 26-41 at the first order in the heavy mass expansion are the following, where $\bar m^2=m_{12}^2/(\sb\cb)$:
\bear
\amp^{\rm 2HDM}_{26}\vert_{\rm heavy} &=& \frac{3\,\mw^2\vev^{-3}\cba^2(\epsilon_1\cdot\epsilon_2)}{q_{45}^2-\mh^2}\left(\cba^2-\sba^2-2\cot2\beta\,\sba\,\cba\right) \nn\\
&&\times\left(4\left(\bar m^2-\mh^2\right)\cba^2\left(\cot2\beta\,\cba+\sba\right)-2\mh^2\sba\right)\,, \nn\\
\amp^{\rm 2HDM}_{27}\vert_{\rm heavy} &=& -\frac{4\,\mw^4\vev^{-3}\sba\,\cba^2\,\left(2\cba^2-1-2\cot2\beta\,\sba\,\cba\right)}{k_{13}^2-\mw^2} \nn\\
&&\times\Big((\epsilon_1\cdot\epsilon_2)-\mw^{-2}(\epsilon_1\cdot k_{13})(\epsilon_2\cdot k_{13})\Big)\,, \nn\\
\amp^{\rm 2HDM}_{28}\vert_{\rm heavy} &=& \amp^{\rm 2HDM}_{27}\vert_{\rm heavy}(k_3\to k_4)\,, \nn\\
\amp^{\rm 2HDM}_{29}\vert_{\rm heavy} &=& -\frac{2\,\mw^2\vev^{-3}\sba\,\cba^2(\epsilon_1\cdot\epsilon_2)}{q_{12}^2-\mh^2}\Big(2\bar m^2\left(2\cba^2-1-2\cot2\beta\,\sba\,\cba\right) \nn\\
&&+\left(-6\bar m^2+4\mh^2+\mHeavy^2+q_{34}^2\right)\left(2\cba^2-1-2\cot2\beta\,\sba\,\cba\right)^2\Big)\,, \nn\\
\amp^{\rm 2HDM}_{30}\vert_{\rm heavy} &=& \amp^{\rm 2HDM}_{29}\vert_{\rm heavy}(k_4\to k_5)\,, \nn\\
\amp^{\rm 2HDM}_{31}\vert_{\rm heavy} &=& \amp^{\rm 2HDM}_{29}\vert_{\rm heavy}(k_3\to k_5)\,, \nn\\
\amp^{\rm 2HDM}_{32}\vert_{\rm heavy} &=& \amp^{\rm 2HDM}_{27}\vert_{\rm heavy}(k_3\to k_5)\,, \nn\\
\amp^{\rm 2HDM}_{33}\vert_{\rm heavy} &=& \amp^{\rm 2HDM}_{27}\vert_{\rm heavy}(k_1\to k_2)\,, \nn\\
\amp^{\rm 2HDM}_{34}\vert_{\rm heavy} &=& \amp^{\rm 2HDM}_{27}\vert_{\rm heavy}(k_1\to k_2,\,k_3\to k_4)\,, \nn\\
\amp^{\rm 2HDM}_{35}\vert_{\rm heavy} &=& \amp^{\rm 2HDM}_{27}\vert_{\rm heavy}(k_1\to k_2,\,k_3\to k_5)\,, \nn\\
\amp^{\rm 2HDM}_{36}\vert_{\rm heavy} &=& \amp^{\rm 2HDM}_{26}\vert_{\rm heavy}(k_5\to k_3)\,, \nn\\
\amp^{\rm 2HDM}_{37}\vert_{\rm heavy} &=& \amp^{\rm 2HDM}_{26}\vert_{\rm heavy}(k_4\to k_3)\,, \nn\\
\amp^{\rm 2HDM}_{38}\vert_{\rm heavy} &=& 6\,\mw^2\vev^{-3}\sba\,\cba^2\,\left(2\cba^2-1-2\cot2\beta\,\sba\,\cba\right)^2(\epsilon_1\cdot\epsilon_2)\,, \nn\\
\amp^{\rm 2HDM}_{39}\vert_{\rm heavy} &=& -4\,\mw^2\vev^{-3}\sba\,\cba^2\,\left(2\cba^2-1-2\cot2\beta\,\sba\,\cba\right)^2(\epsilon_1\cdot\epsilon_2)\,, \nn\\
\amp^{\rm 2HDM}_{40}\vert_{\rm heavy} &=& \amp^{\rm 2HDM}_{39}\vert_{\rm heavy}\,, \nn\\
\amp^{\rm 2HDM}_{41}\vert_{\rm heavy} &=& \amp^{\rm 2HDM}_{39}\vert_{\rm heavy}\,.
\label{ampWWtohhh-2HDM-heavy}
\eear
Next,  we discuss the comparison with the HEFT amplitudes going through the various HEFT coefficients and taking into account the previous classification in \ref{HEFT-classification}. 

\vspace{0.5cm}
{\bf 1) Coefficient $a$}
\vspace{0.5cm}

The diagrams proportional to $a^3$, $\{10,15,18,20,21,24\}$, have no counterpart among the 2HDM diagrams with heavy Higgs bosons, since these ones reduce at most to topologies with a single $W$ propagator. 
Then each of these HEFT diagrams is matched directly to the 2HDM diagram with the same topology,
\bear
\amp^{\rm HEFT}_{10}&=&\amp^{\rm 2HDM}_{10}\,,\nn\\
\amp^{\rm HEFT}_{15}&=&\amp^{\rm 2HDM}_{15}\,,\nn\\
\amp^{\rm HEFT}_{18}&=&\amp^{\rm 2HDM}_{18}\,,\nn\\
\amp^{\rm HEFT}_{20}&=&\amp^{\rm 2HDM}_{20}\,,\nn\\
\amp^{\rm HEFT}_{21}&=&\amp^{\rm 2HDM}_{21}\,,\nn\\
\amp^{\rm HEFT}_{24}&=&\amp^{\rm 2HDM}_{24}\,,
\eear
which gives the following solution:
\be
a =1-\Delta a= \sba \,.
\label{matching-a}
\ee

\vspace{0.5cm}
{\bf 2) Coefficient $b$}
\vspace{0.5cm}

Once $a$ is known, we use the diagrams proportional to $ab$, $\{2,3,6,7,8,11\}$. 
Here the situation is different, since each HEFT diagram is matched to the sum of two 2HDM diagrams: the one with the same topology and the one with $H$ exchange (diagrams 27, 28 and 32-35), which in the heavy mass limit behaves as an effective $WWhh$ vertex. Thus,
\bear
\amp^{\rm HEFT}_2&=&\amp^{\rm 2HDM}_2+\amp^{\rm 2HDM}_{27}\vert_{\rm heavy}\,,\nn\\
\amp^{\rm HEFT}_3&=&\amp^{\rm 2HDM}_3+\amp^{\rm 2HDM}_{28}\vert_{\rm heavy}\,,\nn\\
\amp^{\rm HEFT}_6&=&\amp^{\rm 2HDM}_{6}+\amp^{\rm 2HDM}_{32}\vert_{\rm heavy}\,,\nn\\
\amp^{\rm HEFT}_7&=&\amp^{\rm 2HDM}_7+\amp^{\rm 2HDM}_{33}\vert_{\rm heavy}\,,\nn\\
\amp^{\rm HEFT}_8&=&\amp^{\rm 2HDM}_8+\amp^{\rm 2HDM}_{34}\vert_{\rm heavy}\,,\nn\\
\amp^{\rm HEFT}_{11}&=&\amp^{\rm 2HDM}_{11}+\amp^{\rm 2HDM}_{35}\vert_{\rm heavy}\,.
\eear
Solving these equations for $b$ we get
\be
b = 1-\Delta b= 1 + \cba^2 (1 - 2\cba^2 + 2\cba\sba\cot2\beta)\,.
\label{matching-b}
\ee

\vspace{0.5cm}
{\bf 3) Coefficient $\kappa_3$}
\vspace{0.5cm}

To determine $\kappa_3$ we take the solution for $b$ in \eqref{matching-b} and the diagrams proportional to $b\kappa_3$, $\{1,12,13\}$. 
Each of them is matched to a 2HDM diagram with the same topology plus the diagram with $H$ exchange (diagrams 26, 36 and 37), which in the heavy mass limit becomes a local $WWhh$ interaction in which one of the two Higgs bosons is virtual and decays into two light Higgs bosons through the $hhh$ vertex:
\bear
\amp^{\rm HEFT}_1&=&\amp^{\rm 2HDM}_1+\amp^{\rm 2HDM}_{26}\vert_{\rm heavy}\,,\nn\\
\amp^{\rm HEFT}_{12}&=&\amp^{\rm 2HDM}_{12}+\amp^{\rm 2HDM}_{36}\vert_{\rm heavy}\,,\nn\\
\amp^{\rm HEFT}_{13}&=&\amp^{\rm 2HDM}_{13}+\amp^{\rm 2HDM}_{37}\vert_{\rm heavy}\,.
\eear
The solution of these equations is
\be
\kappa_3 = 1- \Delta \kappa_3 = \sba(1+2\cba^2) +\cba^2\left( -2\sba\mbarh +2\cba\cot2\beta \left(1-\mbarh\right) \right) \,.
\label{matching-kappa3}
\ee
This coefficient is also present in the groups $a\kappa_3^2$ and $a^2\kappa_3$, which do not receive contributions from heavy Higgs bosons in the large mass limit. 
Therefore, the equalities
\bear
\amp^{\rm HEFT}_{i}&=&\amp^{\rm 2HDM}_{i}\,,\qquad i\in\{5,\,9,\,16\}\,,\nn\\
\amp^{\rm HEFT}_{i}&=&\amp^{\rm 2HDM}_{i}\,,\qquad i\in\{14,\,17,\,19,\,22,\,23,\,25\}\,,
\eear
provide two independent checks of \eqref{matching-kappa3}. 

\vspace{0.5cm}
{\bf 4) Coefficients $\kappa_4$ and $c$}
\vspace{0.5cm}

The remaining HEFT diagrams are the contact one, diagram 0, and the one with a quartic Higgs coupling, diagram 4, whose amplitudes have different dependence on $q_{12}$: the first one is a constant and the second one is proportional to $1/(q_{12}^2-\mh^2)$. 
In the 2HDM, the contributions to these two structures come from diagrams 4, 29, 30 and 31, which have a light Higgs propagator in the $q_{12}$ channel, and from diagrams 38-41, where each heavy Higgs propagator is shrunk to a point. 
The matching equation reads
\be
\amp^{\rm HEFT}_0 + \amp^{\rm HEFT}_4 = \sum_{i \in \{4,\, 29,\, 30,\, 31\}} \amp^{\rm 2HDM}_i\vert_{\rm heavy} + \sum_{j \in \{38,\, 39,\, 40,\, 41\}} \amp^{\rm 2HDM}_j\vert_{\rm heavy}\,.
\ee
The sum over $i$ is not just proportional to $(\epsilon_1\cdot\epsilon_2)/(q_{12}^2-\mh^2)$. 
The numerator of the amplitudes 29, 30 and 31 in \eqref{ampWWtohhh-2HDM-heavy} depends on $q_{34}^2$, $q_{35}^2$ and $q_{45}^2$, whose sum is $q_{12}^2+3\mh^2$ by momentum conservation, and then these diagrams also generate a contribution proportional to $\epsilon_1\cdot\epsilon_2$ without propagator. 
This contribution has the same form as the contact amplitude of diagram 0, which is the reason why $\kappa_4$ and $c$ cannot be determined independently. 
Since the two structures are independent functions of $q_{12}^2$, we can equate their coefficients separately. 
The propagator structure involves only $\kappa_4$, and it gives
\bear
\kappa_4 =1-\Delta \kappa_4 &=& 1 +\frac{\cba^2}{3}\left( -7+64\cba^2-76\cba^4 +12(1-6\cba^2+6\cba^4)\mbarh \right.  \nn\\
&&\left. \hspace{7mm}+4\cba\sba\cot 2\beta \left(-13+38\cba^2-3(-5+12\cba^2)\mbarh\right) \right.  \nn\\
&&\left. \hspace{7mm}+4\cba^2\cot^2 2\beta \left(3\cba^2-16\sba^2+3(-1+6\sba^2)\mbarh\right) \right)\,.
\label{matching-kappa4}
\eear
In this equation the terms proportional to $\mHeavy^2$ from diagram 4 and from diagrams 29-31 cancel each other, as anticipated above. 
Once $\kappa_4$ is known, the contact structure fixes $c$:
\be
c=-\frac{4}{3}\,\cba^2\sba\left(2\cba^2-1-2\cba\sba\cot2\beta\right)^2\,.
\label{matching-c}
\ee
It is worth noticing that $c$ is a genuine coefficient  of triple Higgs production and cannot be  determined  from the $WW\to hh$ amplitude at tree level since $c$ does not participate in this amplitude.  We also notice from the above result in \eqref{matching-c} that this $c$ coefficient  is not zero in the 2HDM,  unless the alignment limit is considered or the combination $2\cba^2-1-2\cba\sba\cot2\beta$ vanishes. 

\subsection{HEFT coefficients from non-decoupling heavy Higgs bosons}
We collect here the results of the previous matching equations, written as the departures of the HEFT coefficients from their SM values, 
\bear
\Delta a\vert_{\rm 2HDM} &=& 1 -\sba \,, \nn\\
\Delta b\vert_{\rm 2HDM} &=& - \cba^2 (1 - 2\cba^2 + 2\cba\sba\cot2\beta) \,, \nn\\
\Delta \kappa_3\vert_{\rm 2HDM}   &= & 1- \sba(1+2\cba^2) -\cba^2\left( -2\sba\mbarh +2\cba\cot2\beta \left(1-\mbarh\right) \right) \,, \nn\\
\Delta \kappa_4\vert_{\rm 2HDM}  &=& -\frac{\cba^2}{3}\left( -7+64\cba^2-76\cba^4 +12\left(1-6\cba^2+6\cba^4\right)\mbarh \right.  \nn\\
&&\left. \hspace{7mm}+4\cba\sba\cot 2\beta \left(-13+38\cba^2-3(-5+12\cba^2)\mbarh\right) \right.  \nn\\
&&\left. \hspace{7mm}+4\cba^2\cot^2 2\beta \left(3\cba^2-16\sba^2+3(-1+6\sba^2)\mbarh\right) \right) \,, \nn \\
c\vert_{\rm 2HDM}  &=& -\frac{4}{3}\,\cba^2\sba\left(2\cba^2-1-2\cba\sba\cot2\beta\right)^2 \,.
\label{matching-full}
\eear
These expressions are written in a compact form, using $\sba$, $\sb$, $\cb$ and $\cot2\beta$, which are given in terms of the input parameters in \eqref{trigo}. 
Thus, the five coefficients depend only on $\cba$, $\tanb$, $\mh$ and $m_{12}$, and not on the individual heavy masses, which is the signature of non-decoupling. More precisely, $\Delta a$ depends only on $\cba$, $\Delta b$ and $c$ depend on $\cba$ and $\tanb$, and $\Delta\kappa_3$ and $\Delta\kappa_4$ depend on $\cba$, $\tanb$, $\mh$ and $m_{12}$. 
They are valid for any value $-1\leq\cba\leq 1$, so they describe the general misalignment scenario. 
The coefficients $a$, $b$, $\kappa_3$ and $\kappa_4$ coincide with the ones that were obtained from the matching of double Higgs production in \cite{Arco:2023sac},  while the coefficient $c$ is new. 

In the alignment limit, $\cba=0$, all the LO-HEFT coefficients take their SM values,  as expected:
\bear
\Delta a\vert_{\rm 2HDM}^{\rm al} &=& 0 \,, \nn\\
\Delta b\vert_{\rm 2HDM}^{\rm al} &=&0 \,, \nn\\
\Delta \kappa_3\vert_{\rm 2HDM}^{\rm al}   &= & 0 \,, \nn\\ 
\Delta \kappa_4\vert_{\rm 2HDM}^{\rm al}  &=& 0 \,, \nn\\
c\vert_{\rm 2HDM}^{\rm al}  &=& 0 \,.
\label{matching-align}
\eear

For quasi-alignment, $|\cba|\ll 1$, the results can be simplified by expanding in powers of $\cba$. 
The leading terms of the five coefficients are of ${\cal O}(\cba^2)$, and they read (quasi-alignment, qal) as follows:
\bear
\Delta a\vert_{\rm 2HDM}^{\rm qal} &=&  \frac{\cba^2}{2} \,,  \nn\\
\Delta b\vert_{\rm 2HDM}^{\rm qal} &=& -\cba^2 \,, \nn\\
\Delta \kappa_3\vert_{\rm 2HDM}^{\rm qal}   &= & -\cba^2\left(\frac{3}{2}-2\frac{m_{12}^2}{\mh^2}\frac{1+\tanb^2}{\tanb}\right) \,, \nn\\ 
\Delta \kappa_4\vert_{\rm 2HDM}^{\rm qal}  &=& \cba^2\left(\frac{7}{3}-4\frac{m_{12}^2}{\mh^2}\frac{1+\tanb^2}{\tanb}\right) \,, \nn\\
c\vert_{\rm 2HDM}^{\rm qal}  &=& -\frac{4}{3}\cba^2 \,.
\label{matching-quasialign}
\eear
In this limit the coefficients are correlated and several simple relations among them can be found. The novelty that we find here is a correlation among the new coefficient $c$, $\Delta a$ and $\Delta b$. From \eqref{matching-quasialign} we find that, in the quasi-alignment limit,
\bear
c\vert_{\rm 2HDM}^{\rm qal} &=& -\frac{8}{3}\Delta a\vert_{\rm 2HDM}^{\rm qal} = \frac{4}{3}\Delta b\vert_{\rm 2HDM}^{\rm qal}\,.
\label{correlation-quasialign5}
\eear
In summary,  in this section we have provided the complete analytical solution to the matching equation  between the HEFT  and the full 2HDM amplitudes for  $WW \to hhh$.  This analytical solution is collected in \eqref{matching-full}.  It  provides the set of LO HEFT parameters,   $\Delta a$,   $\Delta b$,  $\Delta \kappa_3$, $\Delta \kappa_4$,  and the new parameter $c$ that contain the 
most relevant non-decoupling effects from the 2HDM heavy Higgs bosons which are present outside the alignment limit,  namely,  whenever $\cba \neq 0$.  
%%%%%%%%%%%%%%%%%%%%%%%%%%%%%%%
\section{Numerical Results}
\label{section-numerical}
 \begin{figure}[!t]
 %   \centering
 \hspace{-0.2cm}
        \includegraphics[scale=0.37]{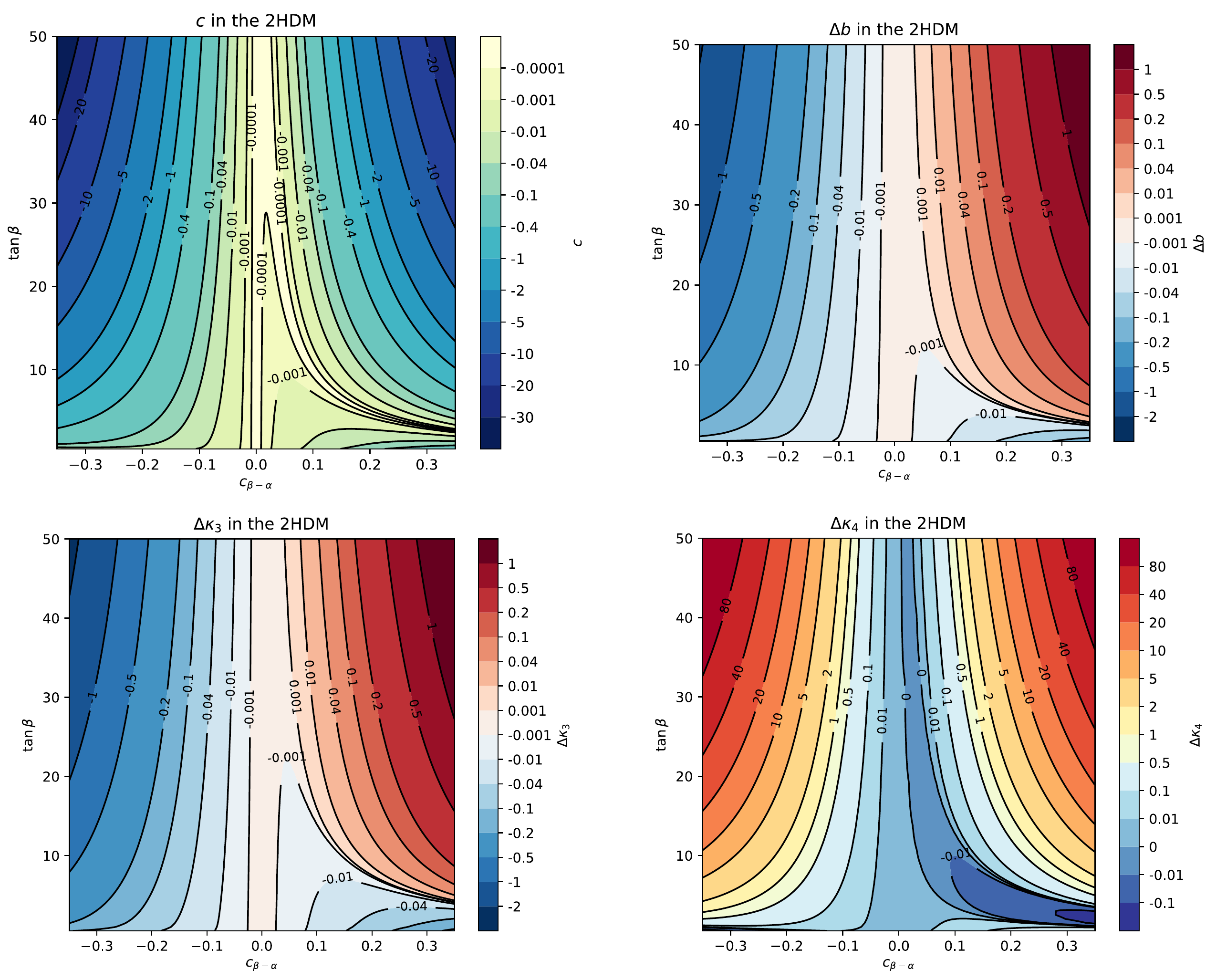}
    \caption{Numerical values of the coefficients when the HEFT is matched to the 2HDM.  The contourlines  show the coefficient values in the 
    $(\cba,\tan \beta)$ plane that are obtained  using the formulas in \eqref{matching-full}: $c$ in the upper left plot,  $\Delta b$ in the upper right plot,  $\Delta \kappa_3$ in the lower left plot, $\Delta \kappa_4$ in the lower right plot.  In the plots of $\Delta\kappa_3$ and $\Delta\kappa_4$ the value  $m_{12}=0$ is set. }
    \label{coeffs-matched}
    \end{figure}
    
In this section we present the numerical results,  first for the  most relevant HEFT coefficients parametrizing the non-decoupling effects from the heavy Higgs bosons of the 2HDM,  and second for the unpolarized cross section $\sigma(WW \to hhh)$ predicted within the HEFT using those coefficients.  The main purpose of this numerical analysis is testing the goodness of these HEFT-matched predictions when compared to the predictions of the full 2HDM.  For the numerical evaluation of the HEFT coefficients we use our python code and for the computations of the cross section we use 
our own implementation of the HEFT and 2HDM models for MadGraph5 (MG5~\cite{Alwall:2014hca}).

In \figref{coeffs-matched} we show the numerical values of the HEFT coefficients,  $c$, $\Delta b$,  $\Delta \kappa_3$ and $\Delta \kappa_4$   in the  $(\cba,\tan \beta)$ plane that are obtained using the formulas in \eqref{matching-full}.  For simplicity,   we set here $m_{12}=0$.  

First,  our numerical estimates in this figure  of $\Delta b$,  $\Delta \kappa_3$ and $\Delta \kappa_4$,  which we show for completeness,  are equivalent to the ones in \cite{Arco:2023sac} since,  as we already said,  the  analytical formulas used for these plots  are coincident.  The case  $m_{12} \neq 0$ is not studied here,  but we refer the reader to  \cite{Arco:2023sac} where this case was also studied.  On the other hand, 
the numerical study of $c$ is new here,   since this coefficient is genuine of triple Higgs production and  does not enter in double Higgs production.  

Second, when comparing the numerical size of the various coefficients,  $c$, $\Delta b$, $\Delta \kappa_3$ and $\Delta \kappa_4$ in this \figref{coeffs-matched},   we see the following features: 1) all them converge to zero value in the alignment limit, i.e.  for $\cba=0$,  as expected; 2) The larger values of the coefficients are reached at the larger values of  $|\cba|$ and $\tb$ in these plots.  The two upper corners in these plots then define the maximum size of the non-decoupling effects we are looking for;  3) The coefficients reaching the largest values are $c$ (negative) and $\Delta \kappa_4$ (positive).   In the case of $c$,  large values with $|c|>1$ are got for $\tan \beta >10$ and for $\cba \neq 0$ distant from the central region.  These large values of $c$ have important implications for the cross section of triple Higgs production as we will comment next.

 \begin{figure}[!t]
%    \centering
 \hspace{-0.5cm}
        \includegraphics[scale=0.35]{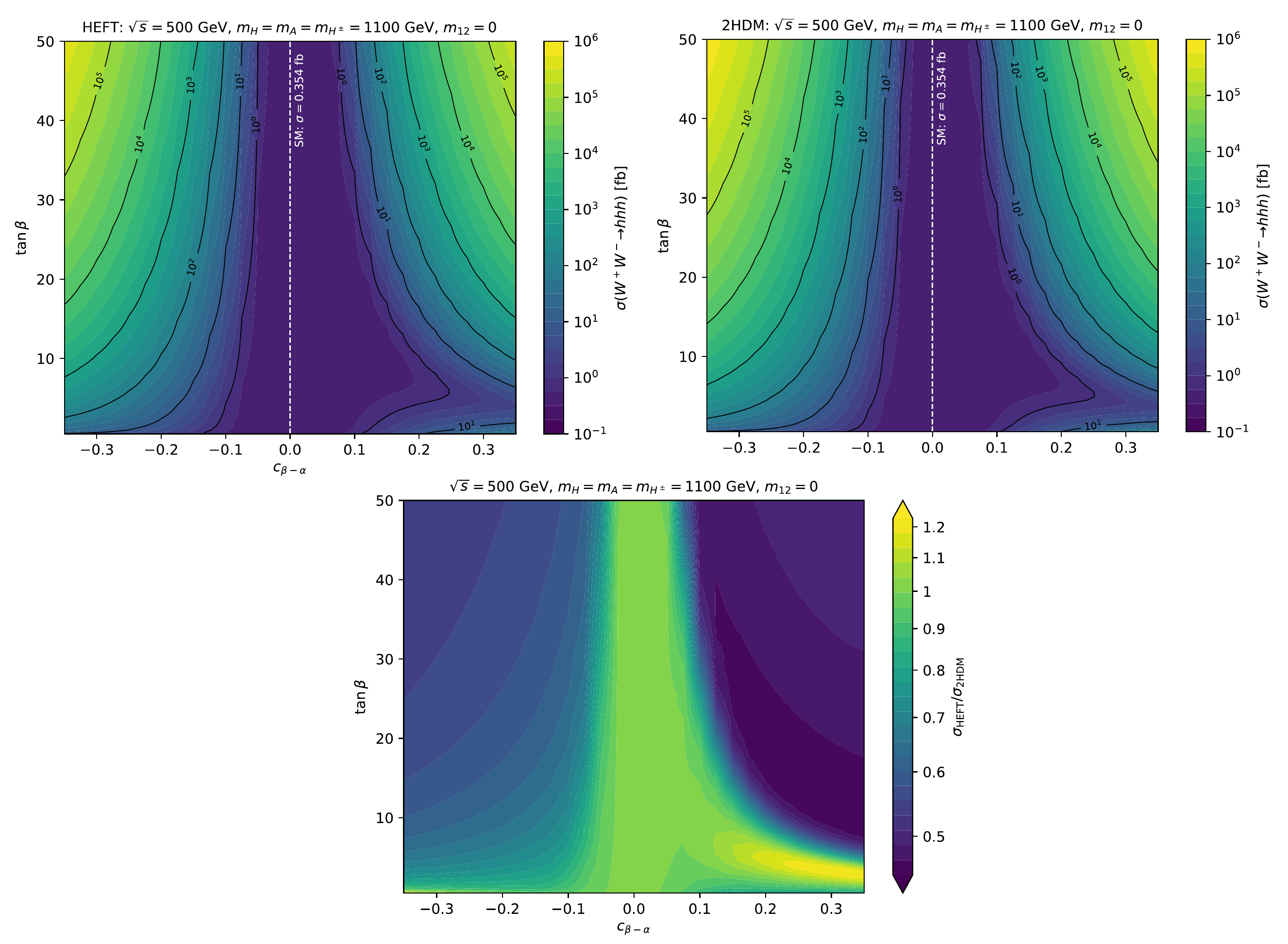}
    \caption{ Predictions of the cross section $\sigma(WW \to hhh)$(fb)  in the $(\cba,\tb)$ plane for a center of mass energy of $\sqrt{s}=500$ GeV.  Left upper plot: results from the HEFT  matched to the 2HDM,  obtained by using the coefficient values of \eqref{matching-full}.  Right upper plot: results from the full 2HDM for $m_{\rm heavy}=m_H=m_A=m_{H^\pm}=1100$ GeV.  Lower plot: Ratio of the two results $\sigma_{\rm HEFT-matched}/\sigma_{\rm 2HDM}$.  In all plots we set $m_{12}=0$. }
    \label{xsection500}
    \end{figure} 

In \figref{xsection500} we show the prediction of the triple Higgs production cross section,  $\sigma(WW \to hhh)$,  from the HEFT when it is matched to the 2HDM by using the values of the coefficients in \eqref{matching-full},  and we also compare it with the prediction from the full 2HDM.  We set here the center of mass energy to $\sqrt{s}=500$ GeV and show the two predictions in the $(\cba, \tan \beta)$ plane:  from the HEFT-matched (upper left plot) which is independent of $m_{\rm heavy}$,  and  from the full 2HDM with the heavy mass set to  $m_{\rm heavy}=m_H=m_A=m_{H^\pm}=1100$ GeV  (upper right plot.)  The ratio of these two cross sections,  $\sigma_{\rm HEFT-matched}/\sigma_{\rm 2HDM}$ is shown in the lower plot.  We see clearly a similar pattern  in the two upper plots,  meaning that the behaviour of the two predictions with these two input parameters $\cba$ and $\tan \beta$ are very close to each other.  The size of the cross section (also similar in the two plots) can depart from the SM value (0.354 fb) in several orders of magnitude,  reaching the largest values 
($> 10^5$ fb) in the upper corners of these plots.  The lower plot shows that the ratio of the two predictions is close to one in the central strip of the plot,  indicating that the approximation provided by the HEFT-matched to the 2HDM works well in that region.   Departures of this ratio from 1 that are larger than  $20\%$  only occur for the larger values of $|\cba|>0.1$ in this plot.

  \begin{figure}[!t]
 %   \centering
 \hspace{-0.5cm}
        \includegraphics[scale=0.33]{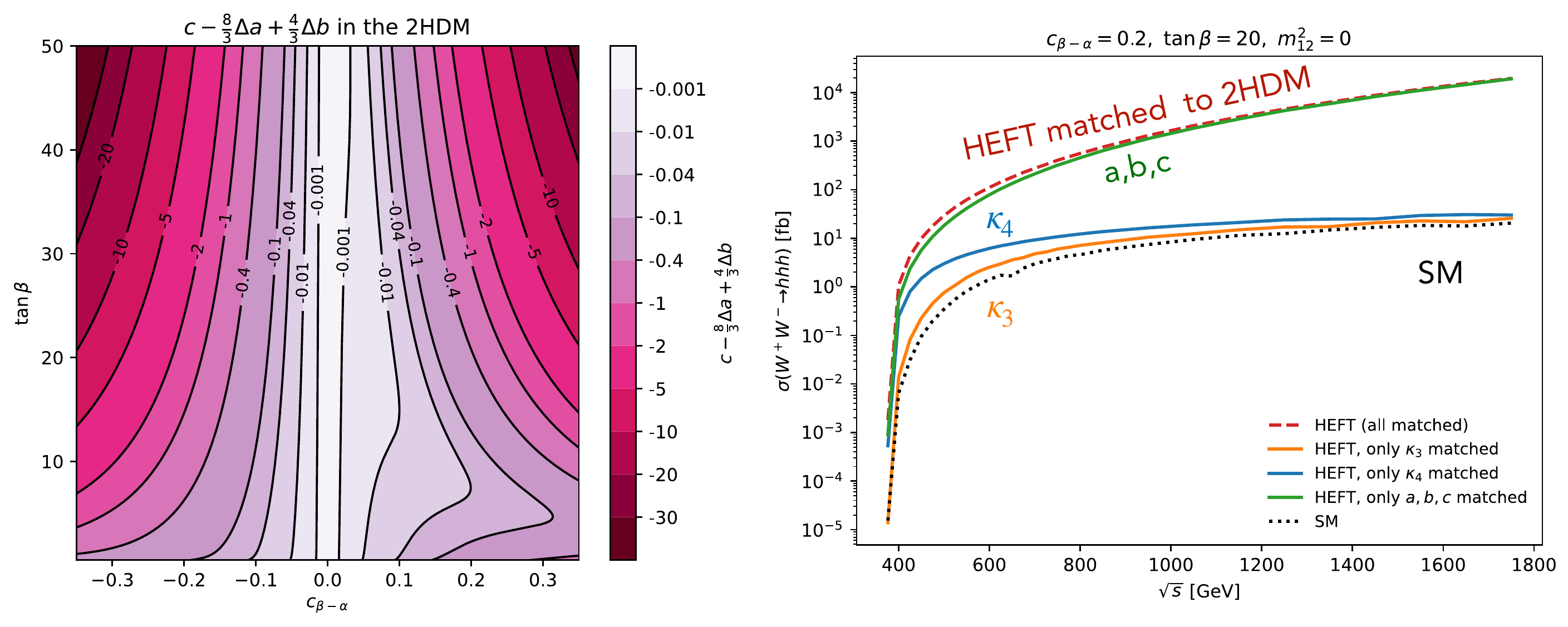}
    \caption{Relevance of the combination $(c-\frac{8}{3} \Delta a+\frac{4}{3} \Delta b)$.  Left plot: Numerical results for $(c-\frac{8}{3} \Delta a+\frac{4}{3} \Delta b)$ in the  $(\cba,\tan \beta)$ plane, when the HEFT is matched to the 2HDM by \eqref{matching-full}.  Right plot:  preditions of $\sigma(WW \to hhh)$(fb) with the HEFT-matched  versus $\sqrt{s}(\rm{GeV})$.  The input parameters are set to $\cba=0.2$,  $\tb =20$,  $m_{12}=0$.  The dashed red line is the total HEFT result.  The dotted black line is the SM result.  The distance HEFT to SM indicates non-decoupling effects. The green line is the contribution from just $a$, $b$ and $c$ (i.e.  setting $\kappa_4=\kappa_3=1$).  The blue line is the contribution from just $\kappa_4$ (i.e setting $a=b=\kappa_3=1$, $c=0$).  The orange line is the 
    contribution from just $\kappa_3$ (i.e setting $a=b=\kappa_4=1$, $c=0$).}
    \label{role-of-abc}
    \end{figure}     
 The relevance of the various HEFT coefficients participating in $\sigma(WW \to hhh)$ is displayed in \figref{role-of-abc}.  The plot in the right shows separately the various contributions to $\sigma(WW \to hhh)$ within HEFT-matched to 2HDM  as a function of the center of mass energy,  $\sqrt{s}$.  Here we set,  $\cba=0.2$ $\tan\beta=20$ and $m_{12}=0$.  Recall that this prediction is independent of $m_{\rm heavy}$.  The dashed line in red is the total HEFT-matched prediction.  The green line is the contribution to the cross section from just $a=1-\Delta a$,  $b=1-\Delta b$,  and $c$ set to the values in  \eqref{matching-full}, and the others set to $\kappa_3=\kappa_4=1$.  The line in blue is the contribution from $\kappa_4=1-\Delta \kappa_4$ set to the value in \eqref{matching-full} and the others set to $a=b=\kappa_3=1$ and $c=0$. The line in orange is the contribution from $\kappa_3=1-\Delta \kappa_3$ set to the value in \eqref{matching-full} and the others set to $a=b=\kappa_4=1$ and $c=0$.  The dotted black line is the SM prediction, included for comparison.   In this plot, we see that both contributions from $\kappa_3$ and $\kappa_4$ stay close to the SM line leading to a nearly flat behaviour with energy at large $\sqrt{s}$.  In contrast,  the contributions from $a$, $b$, and $c$ are the most relevant ones and give the dominant contribution to the total HEFT-matched cross section.  Both lines, the red and the green, show a similar behaviour with $\sqrt{s}$, and  their departure from the SM result also grows similarly with energy,  which is the typical behaviour expected from the EFT approach.  Indeed,  this departure can be quantified with the particular combination of coefficients given by  $c+\frac{4}{3}a(a^2-b)$,  as shown in \cite{Domenech:2025gmn}.  This feature can be demonstrated analytically by selecting the dominant helicity channel coming from the initial longitudinal $W$'s and analyzing the corresponding amplitude at $\sqrt{s} >>m_W, m_h$,  which leads to \cite{Domenech:2025gmn}\footnote{we have corrected here a typo in Eq.  (6.7) of this PRD paper,  as it has been done by the authors themselves in the v3 of the arXive}:
 \begin{eqnarray}
 {\cal A}^{\rm HEFT} (W_L W_L \to hhh) &  \simeq & 3 \frac{s}{v^3} \Big (c+\frac{4}{3}a(a^2-b) \Big ) \nn \\
 & \simeq & 3 \frac{s}{v^3} \Big (c -\frac{8}{3} \Delta a + \frac{4}{3} \Delta b \Big ).
 \end{eqnarray}
 Notice that the value of this combination,  $(c -\frac{8}{3} \Delta a + \frac{4}{3} \Delta b )$,   is non zero outside the alignment limit,   and therefore it represents the most relevant non-decoupling effects from the heavy modes leaving important  imprints in the cross section, which indeed grow in size with $\sqrt{s}$.  Notice also that,  in the quasi-alignment limit,  \eqref{matching-quasialign} and \eqref{correlation-quasialign5},   this combination gives  approximately $3c$,  hence the importance of the $c$ parameter which dominates the rates of the longitudinal $W_L$ modes. 
 \begin{figure}[!t]
 %   \centering
  \hspace{-0.5cm}
        \includegraphics[scale=0.37]{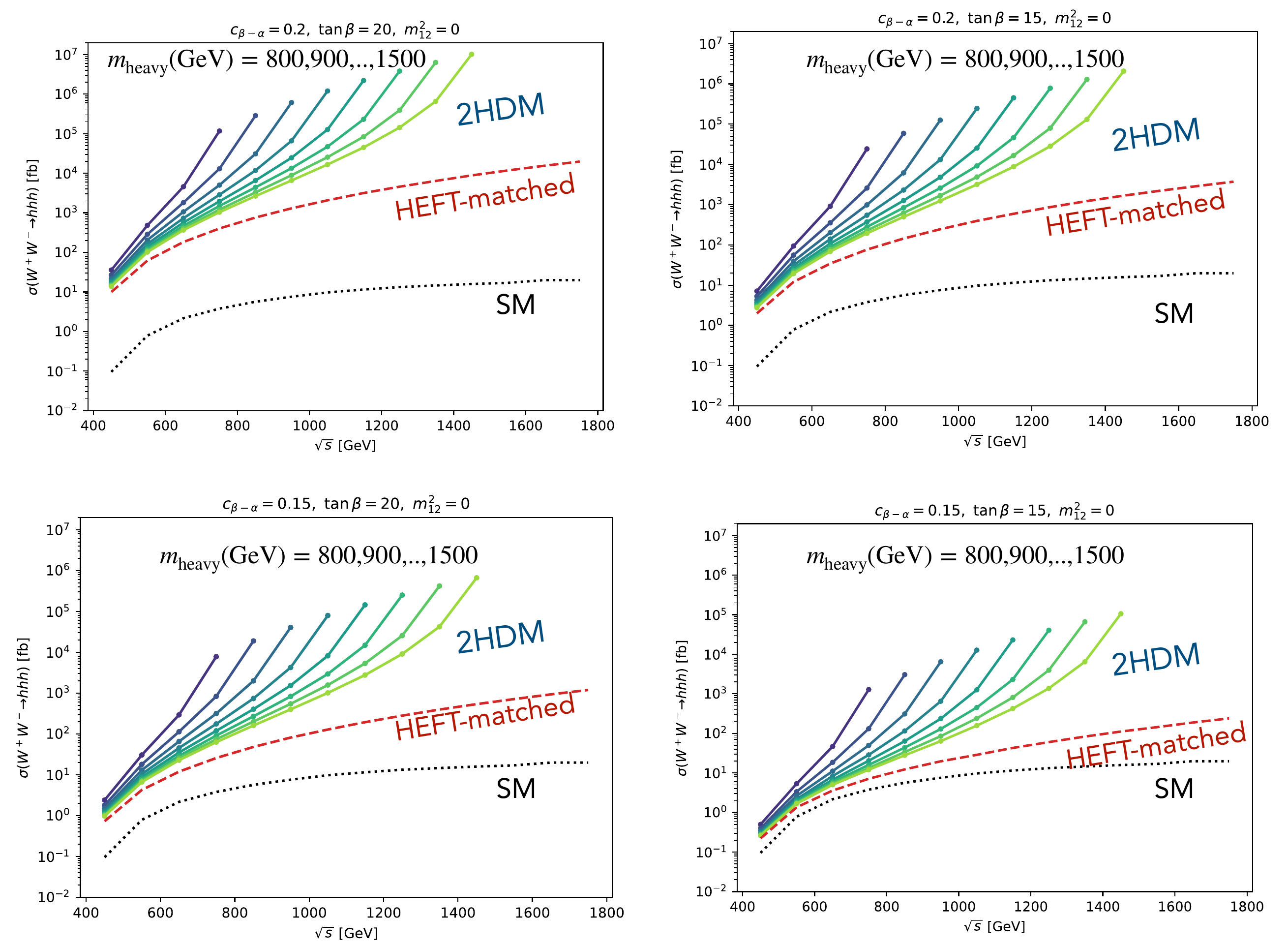}
    \caption{Comparison between HEFT-matched to 2HDM and the full 2HDM.   The  plots show the predictions of $\sigma(WW \to hhh)$(fb) as a function of the center of mass energy $\sqrt{s}({\rm GeV})$.  The dashed red line is the result from the HEFT-matched to 2HDM.  The 8 solid lines (above the dashed red line) colored from blue to green show the predictions from the full 2HDM for 8 values of the heavy mass,  $m_{\rm heavy}({\rm GeV)}=800, 900,...1500$.  The parameter 
    $m_{12}$ is set to zero in all plots.  The other input parameters are set to: $\cba= 0.2$, $\tb= 20$ (upper left plot);  $\cba= 0.2$, $\tb= 15$ (upper right plot); $\cba= 0.15$, $\tb= 20$ (lower left plot); $\cba= 0.15$, $\tb= 15$ (lower right plot).  The SM prediction (dotted black line)  is included in all plots for comparison. }
    \label{sigma-versus-sqrts}
    \end{figure}   
The important role of this particular combination of HEFT coefficients is also shown in the left plot of \figref{role-of-abc}.  This plot displays the values of $(c -\frac{8}{3} \Delta a + \frac{4}{3} \Delta b )$ in the $(\cba, \tan \beta)$ plane,  when the HEFT is matched to the 2HDM by using \eqref{matching-full}.  Again,  the numerical value of this combination goes to zero in the alignment limit and gets the largest  values in the upper corners of this plot.  The most important feature that we learn from this plot, is that this is indeed the proper combination of coefficients that manifests  a similar pattern in  the $(\cba, \tan \beta)$ plane, as that of the total cross section shown in \figref{xsection500}.  Thus,  the behaviour of $\sigma(WW \to hhh)$ with these two parameters $\cba$ and $\tan \beta$ is better explained in terms of this particular combination of $a$,$b$ and $c$ rather than from the behaviour of the separate parameters shown in \figref{coeffs-matched}.
 
 Finally, to quantify the convergence of the full 2HDM result to the HEFT-matched result we display in Fig. \ref{sigma-versus-sqrts} their corresponding cross section predictions verus $\sqrt{s}$,  and we also compare them with the SM predictions.  We choose various points of the $(\cba, \tan \beta)$ plane as illustrative examples: $(0.2, 20)$ (upper left plot),  $(0.2,15)$ (upper right plot),  $(0.15,20)$ (lower left plot) and $(0.15,15)$ (lower right plot).  In all plots we set $m_{12}=0$.  The red dashed line is the HEFT-matched prediction,  the dotted black line is the SM prediction and the eight lines from blue to green color are the predictions from the full 2HDM for eight choices of the heavy mass $m_{\rm heavy}$(GeV)  from 800 to 1500  with increments of  100 GeV.   Notice that we only show the 2HDM predictions for energies below the value of the corresponding $m_{\rm heavy}$(GeV) in order to preserve 
the hierarchy $\sqrt{s} <  m_{\rm heavy}$ which is needed for the heavy mass expansion to make sense.  The main features that we learn from these plots are the following.  The convergence of the full 2HDM result to the HEFT-matched result is clear since the green lines with the heavier  $m_{\rm heavy}$(GeV) values approach the HEFT red line,  and this happens in all plots, i.e for all the studied values of  $(\cba, \tan \beta)$.   Thus, the HEFT-matched  prediction is representing well,  qualitatively and quantitatively,  the non-decoupling effects of the heavy modes.   The gap between the HEFT and the SM  gives precisely the size of this non-decoupling effect,  which is clearly larger for the larger energies. 
We also see in these plots that this gap is reduced in size,  for smaller values of $\cba$ and $\tan \beta$.  Indeed, we have checked that the lines from the 2HDM and the HEFT-matched lie down on top of the SM line when approaching to the alignment limit,  as expected.

%%%%%%%%%%%%%%%%%%%%%%%%%%%%%%%
\section{Conclusions}
\label{section-conclus}
In the present work we have studied in full detail the comparison of the HEFT predictions with the 2HDM predictions for   triple Higgs production via $WW \to hhh$ scattering,   under the hypothesis of very heavy Higgs bosons  $H$,  $A$,  and $H^\pm$,  and assuming the lightest one $h$ has the mass observed in the experiments,  $m_h \simeq 125$ GeV. 
We have solved the matching equations relating the two amplitudes,  from the HEFT and from the full 2HDM in the heavy mass limit,  
$m_{\rm heavy} \gg m_{\rm EW}$, and we have found the values of the LO-HEFT coefficients,  $\Delta a$, $ \Delta b$,  $c$, $\kappa_3$, $\kappa_4$ as a function of the 2HDM input parameters.  The analytical solution for these coefficients is summarized in \eqref{matching-full}.   We have found that the solutions for $\Delta a$, $ \Delta b$, $\Delta \kappa_3$ and $\Delta \kappa_4$ coincide with the solutions for the double Higgs production case found previously in \cite{Arco:2023sac}.  The solution of the amplitude matching equation for the contact parameter $c$ is new,  since it is a genuine coefficient  of triple Higgs production and does not participate in double Higgs production processes.  It is also worth mentioning that these same solutions in \eqref{matching-full} have also been found by the different functional method to integrate out the heavy modes  in \cite{Buchalla:2023hqk}.  This comparison is an interesting check of our analytical results from the scattering amplitudes for triple Higgs production where we use a diagrammatic method.  Finally, we have also checked these analytical solutions with the MATCHA code presented recently in \cite{Gomez-Ambrosio:2026fvq}.

The solutions found for the HEFT-matched coefficients summarize the leading effects at low energies from the 2HDM heavy Higgs bosons and have important phenomenological consequences. They represent the non-decoupling effects of these heavy modes since they leave contributions to the scattering amplitudes (and therefore to the observable cross section) that do not vanish as inverse powers of the heavy masses.  The relevance of these  non-decoupling contributions to $\sigma(WW \to hhh)$ has also been analized numerically in this work.  From this numerical analysis we have concluded,  first that the approximation provided by the HEFT-matched to the 2HDM works well,  whenever $m_{\rm heavy}$ is above the energy $\sqrt{s}$  of the $WW \to hhh$ process, and whenever the value of $\cba$ is not too far from the alignment limit.  The quasi-alignment region,  defined as $\cba \neq 0$ but still small (say $\leq 0.2$) provides the optimal region where to apply our results for the matched coefficients. 

The numerical results of the cross section,  $\sigma(WW \to hhh)$,  for the HEFT-matched to the 2HDM also show that the most relevant coefficients are
$a$, $b$ and $c$ and confirm that indeed the particular combination $(c -\frac{8}{3} \Delta a + \frac{4}{3} \Delta b )$ is the one driving both the behaviour with the total energy $\sqrt{s}$ and the behaviour with the two input parameters $\cba$ and $\tan \beta$.  The role of $\Delta \kappa_3$ and $\Delta\kappa_4$ is clearly subleading at these high energies around the TeV scale.  Finally,  the separation with respect to the SM prediction is fully understood in terms of these parameters.  Therefore,  the non-decoupling contributions found here characterize optimally the new physics departures with respect to the SM coming from the 2HDM heavy modes , $H$, $A$ and $H^{\pm}$ which can be used for the searches of new physics at high energy colliders, even if the energy is not enough to produce these heavy modes on-shell.

%%%%%%%%%%%%%%%%%%%%%
\section*{Acknowledgments}

We  wish to warmly thank Roberto Morales for his kind help with the model files for the computation within the HEFT and the 2HDM.  We also thank him  and Francisco Arco for their contributions to the previous referent work where the matching procedure was set for the double Higgs production case.  This  previous work has also served us for comparison with the present case of triple Higgs production. 
We acknowledge partial financial support by the Spanish Research Agency (Agencia Estatal de Investigación) through the grant  with reference number PID2025-172338NB-I00 funded by MICIU/AEI/10.13039/501100011033 and by ERDF, EU.
% PID2022-137127NB-I00 
%funded by MCIN/AEI/10.13039/501100011033/ FEDER, UE.  
We also acknowledge financial support from the  grant IFT Centro de Excelencia Severo Ochoa with reference number CEX2020-001007-S
 funded by MCIN/AEI/10.13039/501100011033,  and from the European Union’s Horizon 2020 research and innovation programme under the Marie Sklodowska-Curie grant agreement No 860881-HIDDeN.  
The work of D.D.  is also supported by the Spanish Ministry of Science and Innovation via an FPU grant No FPU22/03485. 

\newpage

%%% la bibliografia
\bibliography{BDGHS.bib}

\end{document}